\documentclass[12pt,a4paper]{article}
\usepackage{amsmath,amsfonts}
\usepackage{algorithm}
\usepackage{array}
\usepackage[caption=false,font=footnotesize]{subfig}
\usepackage{textcomp}
\usepackage{stfloats}
\usepackage{url}
\usepackage{verbatim}
\usepackage{graphicx}
\usepackage{cite}
\usepackage{hyperref}
\usepackage{algpseudocode}
\usepackage{xcolor}
\usepackage{geometry}
\usepackage{setspace}
\begin{document}

\title{\textbf{Channel Knowledge Map Construction via Physics-Inspired Diffusion Model Without Prior Observations}}

\author{Yunzhe~Zhu, Xuewen~Liao, Zhenzhen~Gao, Linzhou~Zeng, Yong~Zeng}


\date{}
\maketitle

\begin{abstract}
  The ability to construct channel knowledge map (CKM) with high precision is essential for environment awareness in 6G wireless systems. However, most existing CKM construction methods formulate the task as an image super-resolution or generation problem, thereby employing models originally developed for computer vision. As a result, the generated CKMs often fail to capture the underlying physical characteristics of wireless propagation. In this paper, considering that acquiring channel observations incurs non-negligible time and cost, we focus on constructing CKM for large-scale fading scenarios without relying on prior observations, and we design three physics-based constraints to characterize the spatial distribution patterns of large-scale fading. By integrating these physical constraints with state-of-the-art diffusion model that possesses superior generative capability, a physics-inspired diffusion model for CKM construction is proposed. Following this motivation, we derive the loss function of the diffusion model augmented with physics-based constraint terms and further design the training and generation framework for the proposed physics-inspired CKM generation diffusion model. Extensive experiments show that our approach outperforms all existing methods in terms of construction accuracy. Moreover, the proposed model provides a unified and effective framework with strong potential for generating diverse, accurate, and physically consistent CKM.
\end{abstract}


\section{Introduction}
In sixth-generation (6G) communication systems, the concept of sensing is expected to evolve from spectrum-level sensing in cognitive radio to environment-level sensing in integrated sensing and communication (ISAC) systems, and further to semantic-level sensing for intelligent network resource optimization. Since electromagnetic waves experience attenuation, reflection, scattering, and diffusion during propagation, storing channel-related information at each spatial location enables the establishment of a link between the physical environment and the wireless communication system. Such spatially distributed data, referred to as channel knowledge map (CKM)\cite{Zeng2021}, serves as a knowledge-driven representation of the propagation environment. CKM not only facilitates a deeper understanding of environmental information by communication systems but also provides essential support for semantic-level perception required in intelligent network resource management, thereby attracting significant attention in 6G research.

\begin{figure}[!t]
\centering
\includegraphics[width=\columnwidth]{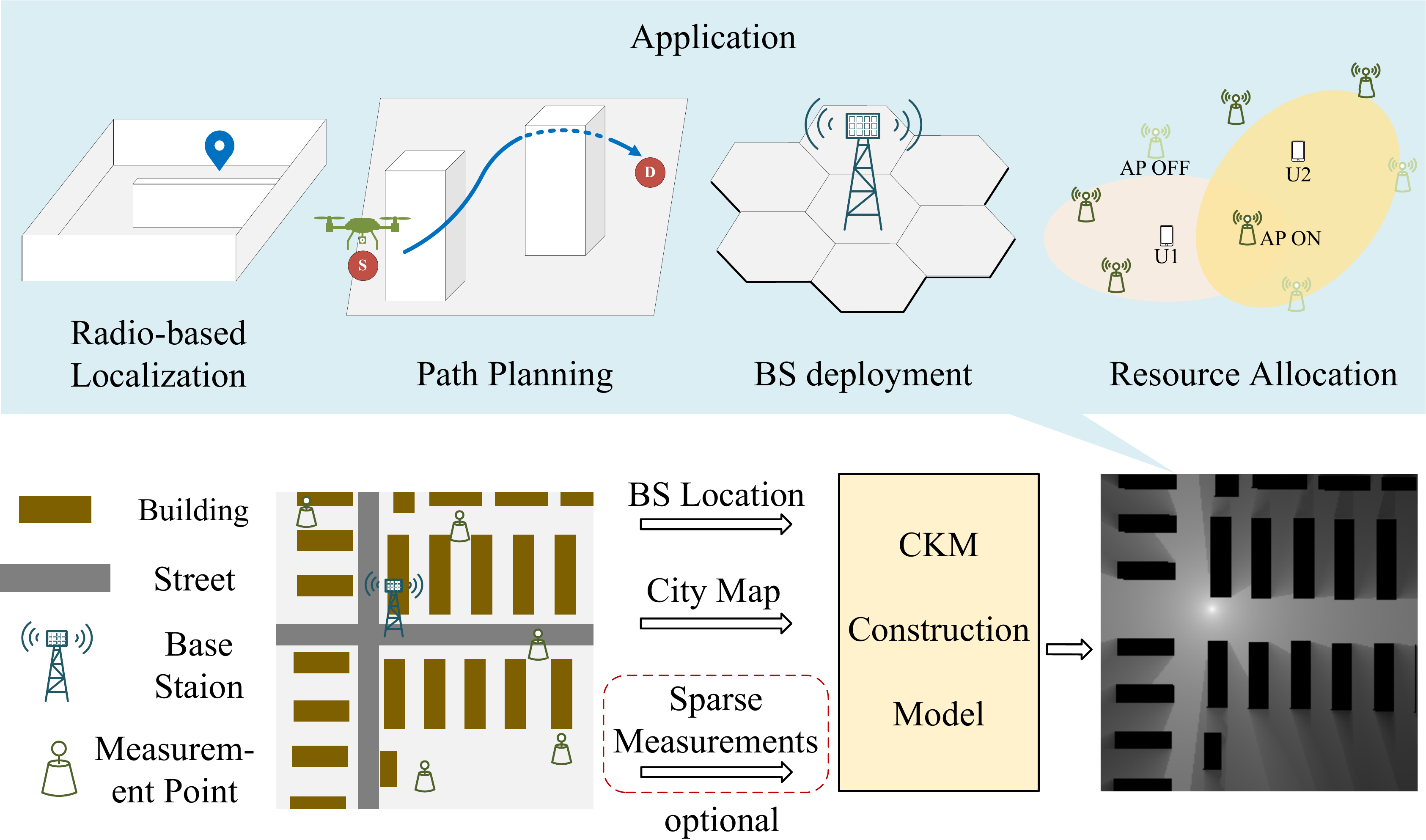}
\caption{Construction and Application of Channel Knowledge Map.}
\label{fig_1}
\end{figure}

CKMs can be categorized into various types according to the channel-related information they represent. Such information may include large-scale fading features such as line-of-sight (LOS) and non-line-of-sight (NLOS) states, path loss, and shadow fading, as well as small-scale fading features such as multipath delay, Doppler power spectrum, and angular power spectrum. In this work, we investigate the channel gain map (CGM), a specific form of CKM that characterizes the path loss value at each spatial position within a region. In some studies, this type of map is also referred to as Radio Map. A more comprehensive discussion of the concepts of Radio Map and CKM can be found in surveys\cite{Romero2022} and\cite{Zeng2024}. After obtaining the CKM, its major applications in wireless communication systems can be categorized into four types, as illustrated in Fig. \ref{fig_1}: base station (BS) placement or deployment, unmanned aerial vehicle (UAV) trajectory planning, user resource allocation, and radio-based localization.

Radio-based localization is one of the most prevalent application areas of CKM, as location information can be inferred from CKM based on the user’s online estimation of channel-related features. The authors in\cite{9031749} and\cite{10025691} respectively propose constructing a received signal strength (RSS) map and an angle–delay domain channel impulse response map for localization. In addition, UAV path planning constitutes another important application area of CKM. As demonstrated in\cite{8525324}, LOS/NLOS state maps can be exploited to jointly optimize UAV trajectories and user scheduling, thereby maximizing the aggregated data throughput of all user nodes within a limited flight time. Furthermore,\cite{9269485} shows that, by constructing a signal-to-interference-plus-noise ratio (SINR) map, the Dijkstra algorithm can be employed to identify the shortest path that consistently satisfies communication quality requirements while avoiding co-channel interference from neighboring terrestrial BS. In addition, the authors of\cite{9354009} employed deep reinforcement learning to plan the trajectory of cellular-connected UAVs, and simultaneously constructing CKM to improve the training efficiency of the learning agent. Meanwhile, existing studies have demonstrated that employing CKM in BS placement or configuration can yield notable performance gains and cost reductions. For example,\cite{11063460} presents a CGM-based scheme that optimizes the antenna downtilt and azimuth angles to maximize cell coverage and \cite{10008712} proposes a CGM-enabled BS deployment method that satisfies multi-BS coverage requirements and quality-of-service constraints. Since CKMs can accurately characterize the spatial properties of wireless channels, their applications can be extended to higher-layer network functionalities such as user resource allocation and scheduling. The authors in\cite{9104036} develop a neural-network–based power control scheme leveraging large-scale fading coefficients. Hence, constructing multi-cell CGMs can provide valuable support for resource allocation and power control in multi-cell massive MIMO systems. Moreover, as shown in\cite{10474197}, CKM can be employed in cell-free massive MIMO (CF-mMIMO) systems to facilitate access-point (AP) switching, thereby mitigating pilot contamination and reducing energy consumption associated with path-loss estimation. Beyond the aforementioned applications, CKM has also attracted increasing attention in a broader range of areas, such as ISAC\cite{10693799}, physical layer security\cite{8815467}, intelligent beamforming\cite{10287775}, and device-to-device communications\cite{9053347}.

Considering that obtaining complete channel information over an entire spatial region purely through measurements is costly, time-consuming, and practically infeasible, research efforts have been devoted to developing efficient CKM construction methods. Depending on the implementation approach, existing CKM construction techniques can be categorized into
four types: model-based methods using statistical channel models, model-based methods using deterministic channel models, spatial interpolation–based methods, and deep learning–based methods. The statistical channel model-based construction methods refer to utilizing existing statistical channel models to compute large-scale or small-scale fading channel information. Commonly used statistical channel models include the log-distance model\cite{5635444} and the Okumura–Hata model\cite{1622772} for large-scale fading computation, as well as Quadriga\cite{6758357} and the 6G PCM (pervasive channel model)\cite{10225614} for small-scale fading modeling. Although this method is simple, it often yields CKMs with limited accuracy. The deterministic channel model–based construction methods obtain channel-related information by performing ray-tracing–based channel simulations. Representative ray-tracing simulators include Wireless InSite\cite{RemcomInc2024}, WinProp\cite{7916282}, and Sionna\cite{Hoydis2022SionnaAO}, among others. Although this approach can achieve high accuracy, it suffers from extremely high computational complexity, making it difficult to meet the real-time requirements of physical-layer algorithms.

The spatial interpolation–based approach constructs CKMs using a limited number of measured channel samples at selected locations. By exploiting the spatial correlation, smoothness, or low-rank characteristics of the wireless channel over the spatial domain, classical interpolation algorithms are applied to estimate the channel information in unmeasured regions. Typical spatial interpolation algorithms used for CKM construction include inverse distance weighting (IDW)\cite{Lu2008}, Kriging\cite{1371308}, radial basis function (RBF) interpolation\cite{Bishop1995}, matrix completion\cite{7472907}, compressed sensing\cite{5352337}, and the expectation–maximization (EM) algorithm\cite{9771802}. Furthermore, the number of samples required to construct a CKM using spatial interpolation methods has been theoretically analyzed in \cite{10530520}. Although the above methods significantly reduce the computational complexity compared with ray-tracing–based approaches, they do not incorporate physical environmental information. Moreover, these spatial interpolation algorithms rely on the assumptions of strong spatial correlation, smoothness, or low-rank characteristics of the channel. In densely built urban areas, where such assumptions are often violated, the accuracy of CKMs obtained through interpolation can degrade considerably.

Deep learning models can effectively establish a mapping from the physical environment map to CKM, thereby enabling more accurate and generalizable CKM construction. The earliest attempts to apply deep learning for CKM construction were inspired by models in the field of computer vision. Specifically, convolutional neural networks based on autoencoder architectures \cite{10682525,9523765}, UNet structures\cite{9354041}, and Laplacian pyramid frameworks\cite{10791446} were introduced to capture the spatial correlations of large-scale fading channels. 

The discriminative models employed in the aforementioned studies typically yield over-smoothed predictions that lack sufficient high-frequency details. However, the spatial distribution of channel-related features in complex environments exhibits strong nonlinearity and fine-grained variations, making it difficult for purely supervised learning to accurately capture such characteristics. In contrast, generative models learn the underlying data distribution and are therefore capable of capturing higher-order statistical characteristics, enabling more accurate prediction of fine-grained spatial variations in channel information. Previous studies\cite{8794603,10130091} have developed conditional GAN (cGAN)–based frameworks to achieve environment-consistent CKM generation. However, the GAN framework involves a minimax optimization objective, which often leads to unstable training and convergence difficulties. In recent years, diffusion models have been recognized as a more powerful class of generative models, providing significantly higher generation quality than GANs. Inspired by their success in other domains, researchers have introduced conditional diffusion frameworks for CKM construction. The studies in\cite{fu2025ckmdiff,10764739} have demonstrated that conditional diffusion models can achieve state-of-the-art CKM prediction accuracy under various input conditions.

However, the aforementioned studies that employ diffusion models for CKM construction directly transfer conditional diffusion frameworks from the field of computer vision. In image generation tasks, various improved conditional diffusion models are designed to produce images that are more realistic and visually appealing. Consequently, their performance is typically evaluated using image quality metrics such as the structural similarity index measure(SSIM), Fréchet inception distance(FID), and learned perceptual image patch similarity. In contrast, the primary objective in CKM construction is to achieve high prediction accuracy rather than visual fidelity. Therefore, conditional diffusion models with high image generation quality are not suitable for direct application to CKM construction. In this paper, motivated by the physical characteristics of wireless signal propagation, we integrate physical prior knowledge with the diffusion model and propose a physics-inspired diffusion-based CKM generation framework. The proposed model not only inherits the advantages of diffusion models in generating high-quality results with richer high-frequency details, but also ensures that the generated CKMs conform to the physical constraints of wireless signal propagation. It is worth noting that several physics-informed methods for constructing CGM have been proposed in\cite{10608081} and\cite{shahid2025reveal}. These approaches introduce physical constraints derived from Maxwell’s equations and incorporate them into the neural network training process as additional loss terms. However, we note that Maxwell’s equations represent the microscopic formulation of wireless signal propagation, as the described electric field variations occur at the wavelength scale. Therefore, they are not well-suited to serve as physical constraints for large-scale fading, which characterizes the macroscopic propagation behavior of wireless signals. Moreover, the boundary conditions of electromagnetic reflections, refractions, and diffractions are highly complex, and existing physics-inspired methods do not impose boundary-condition residuals as training constraints, further reducing the physical fidelity of the CGM construction. 
To address these limitations, this paper heuristically designs three novel physical constraint terms that better characterize the spatial distribution of large-scale fading, and innovatively integrates them into the diffusion model framework.

Based on the above discussions, the main contributions of this paper can be summarized as follows:
\begin{enumerate}
\item{To address the insufficient physical consistency exhibited by existing CGM construction methods, we design three physics-inspired regularization terms, namely the edge loss, the regional propagation loss, and the multi-scale feature loss. These three regularizations respectively reflect distinct physical phenomena exhibited by electromagnetic waves when encountering obstacles, and are therefore closely tied to the propagation environment. This makes them particularly suitable for CGM construction in complex scenarios such as urban canyons and indoor environments.}
\item{We propose integrating the physics-based regularization terms with the diffusion model to generate CGMs. Since the vanilla diffusion model is optimized by minimizing the evidence lower bound (ELBO) derived from variational inference, its optimization objective is not inherently aligned with minimizing the residuals of the physics-inspired terms. To address this inconsistency, we derive the modified objective function of the diffusion model that incorporates the proposed physical regularizations, thereby achieving a seamless fusion of physical prior knowledge and the diffusion generation framework.}
\item{To enable the computation of the physical regularization terms during training, we design a three-stage training strategy together with a corresponding physics-inspired diffusion training framework. This design not only accelerates the training process but also ensures the accurate computation of the physical regularization terms, thereby guaranteeing that the generated CKMs satisfy the intended physical constraints.}
\end{enumerate}

The remainder of this paper is organized as follows. Section II describes the CKM construction problem. Section III introduces several classical diffusion models and their underlying mathematical principles, upon which the proposed model in this paper is developed. Section IV presents the details of the proposed physics-inspired CKM generation diffusion model, including the computation of the three physics-based regularization terms, the derivation of the loss function, and the design of the training and generation framework as well as its algorithmic workflow. Section V provides the experimental results, including comparisons with state-of-the-art CGM construction methods and ablation studies. Finally, Section VI concludes the paper.

\section{CKM Construction Problem Description}
The objective of this paper is to construct a classical type of CKM—the CGM. In the construction of CKM, including the CGM, the map is typically divided into $H \times H$ grids. Consequently, both the ground-truth CGM $\mathbf{x}_{GT}$ and the predicted CGM $\mathbf{x}_{pred}$ are represented as real-valued matrices of size $H \times H$. Accordingly, when external conditional information is required for CGM construction, the input data are also represented as real-valued matrices of the same size. In the transmitter location matrix $\mathbf{L}_{\text{TX}} \in \mathbb{R}^{H \times H}$, only the element corresponding to the transmitter position is set to 1, while all other elements are 0. In the building environment matrix $\mathbf{B} \in \mathbb{R}^{H \times H}$, the elements representing building regions are set to 1, and all others are 0. In the vehicle location matrix $\mathbf{C} \in \mathbb{R}^{H \times H}$, the element corresponding to the vehicle position is set to 1, with all remaining elements being 0. In the sparse observation matrix $\mathbf{S} \in \mathbb{R}^{H \times H}$, the elements at the observation positions contain the actual observed values, whereas all other elements are set to 0.

According to the type of input data, the CGM construction problem can be generally classified into four categories, namely the purely model-driven construction without any prior observations or environmental data, the construction relying solely on sparse observations, the construction based exclusively on propagation environment information, and the joint construction that simultaneously exploits both sparse observations and environmental features. In the absence of any external input, the construction of CGM can only rely on statistical channel models, whose accuracy is inherently limited and thus inadequate for practical applications. When sparse observations are available as external inputs, the construction requires the measured or estimated path loss at locations that are sparsely yet uniformly randomly sampled within the region of interest. However, this CGM construction approach based on sparse observations faces several constraints. First, certain positions within the area are difficult to access. Second, acquiring these sparse observations through channel sounding or channel estimation introduces additional cost and overhead. Finally, when moving obstacles such as vehicles exist within the region, the CGM becomes dynamically varying. In this case, the construction algorithm is required to rapidly generate a CGM snapshot that represents the continuously changing CGM. Nevertheless, measuring or estimating path loss requires extra time, which makes it difficult to meet the requirements of rapid and real-time CGM construction. Building upon the above analysis, constructing the CGM solely based on the propagation environment is considered to be of the highest practical significance and application value. Therefore, this paper investigates the construction of the CGM under given propagation environment conditions. When the environment contains only buildings, the CGM is regarded as a static CGM, and the construction problem can be formulated as follows:
\begin{equation}
\label{deqn_ex1a}
\mathbf{x}_{{pred}}^S=f_\theta\left(\mathbf{L}_{\text{TX}}, \mathbf{B}\right).
\end{equation}
When vehicles are also present in the environment, the map is referred to as a dynamic CGM, and the construction problem can be formulated as follows:
\begin{equation}
\label{deqn_ex2a}
\mathbf{x}_{{pred}}^D=f_\theta\left(\mathbf{L}_{\text{TX}}, \mathbf{B}, \mathbf{C}\right),
\end{equation}
where $f_\theta$ denotes the CGM construction model.

\section{Theoretical Foundations of Diffusion Models}
This section introduces the fundamental concepts of several diffusion models that are utilized in the proposed physics-inspired CKM generation diffusion model.
\subsection{Denoising Diffusion Probabilistic Model}
Inspired by nonequilibrium thermodynamics, Denoising Diffusion Probabilistic Model (DDPM)\cite{ho2020denoising} formulates the forward noise-adding and reverse denoising processes as Markov chains. Therefore, DDPM mainly consists of two components: a forward process that gradually adds noise to the input, and a reverse process that learns how to denoise it.

The forward diffusion process in DDPM is a fixed Markov process, where $\mathbf{x}_{t-1}$ denotes the input data at time step $t-1$, and $\mathbf{x}_t$ represents the corresponding noised data at time step $t$. The noise-adding process from $\mathbf{x}_{t-1}$ to $\mathbf{x}_t$ is modeled as follows:
\begin{equation}
\label{deqn_ex3a}
\mathbf{x}_t=\sqrt{1-\beta_t} \mathbf{x}_{t-1}+\sqrt{\beta_t} \boldsymbol{\epsilon}_t, \boldsymbol{\epsilon}_t \sim \mathcal{N}\left(\mathbf{0}, \mathbf{I}\right),
\end{equation}
where $\beta_t$ is a hyperparameter. The above noise-adding process can be expressed in the form of a marginal distribution as follows:
\begin{equation}
\label{deqn_ex4a}
q(\mathbf{x}_t\mid\mathbf{x}_{t-1})=\mathcal{N}\left(\mathbf{x}_t ; \sqrt{1-\beta_t} \mathbf{x}_{t-1}, \beta_t \mathbf{I}\right).
\end{equation}
Furthermore, the data at any time step can be derived from the original input $\mathbf{x}_0$ through the following expression:
\begin{equation}
\label{deqn_ex5a}
\mathbf{x}_t=\sqrt{\bar{\alpha}_t} \mathbf{x}_0+\sqrt{1-\bar{\alpha}_t} \boldsymbol{\bar{\epsilon}}_t,
\end{equation}
where $\bar{\alpha}_t=\prod_{s=1}^t \alpha_s$, $\alpha_s=1-\beta_s$. Therefore, if the number of time steps $T$ is sufficiently large, the data obtained after $T$ iterative noise-adding steps will follow a standard normal distribution. The core idea of the diffusion model is to learn the true reverse posterior distribution $q(\mathbf{x}_{t-1}\!\mid\!\mathbf{x}_{t})$, enabling the model to generate data whose distribution $p(\mathbf{x}_{0})$ closely matches the data distribution $q(\mathbf{x}_{0})$ of the training set. In practice, the complex distribution $q(\mathbf{x}_{t-1}\!\mid\!\mathbf{x}_{t})$ is often approximated by a Gaussian distribution. However, due to the intractability of directly computing $q(\mathbf{x}_{t-1}\!\mid\!\mathbf{x}_{t})$ using Bayes’ theorem, the posterior distribution $q(\mathbf{x}_{t-1}\!\mid\!\mathbf{x}_{t},\mathbf{x}_{0})$ is usually evaluated instead. Through derivation, the analytical expression of the posterior distribution $q(\mathbf{x}_{t-1}\!\mid\!\mathbf{x}_{t},\mathbf{x}_{0})$ can be obtained as follows:
\begin{equation}
\label{deqn_ex6a}
\begin{aligned}
q(\mathbf{x}_{t-1} \mid \mathbf{x}_t, \mathbf{x}_0) & =\frac{q(\mathbf{x}_t \mid \mathbf{x}_{t-1}, \mathbf{x}_0) q(\mathbf{x}_{t-1} \mid \mathbf{x}_0)}{q(\mathbf{x}_t \mid \mathbf{x}_0)} \propto \mathcal{N}\left(\mathbf{x}_{t-1} ; \tilde{\mu}_t, \tilde{\beta}_t \mathbf{I}\right),
\end{aligned}
\end{equation}
where $\tilde{\mu}_t$ and $\tilde{\beta}_t$ are respectively:
\begin{equation}
\label{deqn_ex7a}
\begin{aligned}
& \tilde{\mu}_t=\frac{\sqrt{\alpha_t}(1-\bar{\alpha}_{t-1}) \mathbf{x}_t+\sqrt{\bar{\alpha}_{t-1}}(1-\alpha_t) \mathbf{x}_0}{1-\bar{\alpha}_t}, \\
& \tilde{\beta}_t=\frac{(1-\alpha_t)\left(1-\bar{\alpha}_{t-1}\right)}{1-\bar{\alpha}_t},
\end{aligned}
\end{equation}
the mean $\tilde{\mu}_t$ can be reformulated into the following form according to Eq.(\ref{deqn_ex5a}):
\begin{equation}
\label{deqn_ex8a}
\tilde{\mu}_t=\frac{1}{\sqrt{\alpha_t}}\left(\mathbf{x}_t-\frac{1-\alpha_t}{\sqrt{1-\bar{\alpha}_t}} \boldsymbol{\bar{\epsilon}}_t\right).
\end{equation}
According to the observation of Eq.(\ref{deqn_ex7a}), the variance of $q(\mathbf{x}_{t-1}\!\mid\!\mathbf{x}_{t},\mathbf{x}_{0})$ is a constant determined by the hyperparameters, while in the mean term Eq.(\ref{deqn_ex8a}) all components except for $\boldsymbol{\bar{\epsilon}}_t$ are also fixed. Therefore, in diffusion models, a neural network is employed to predict $\boldsymbol{\bar{\epsilon}}_t$, which represents the noise introduced during the forward noising process at time step $t$. Consequently, the training objective of the diffusion model is defined as follows:
\begin{equation}
\label{deqn_ex9a}
L=\left\|\boldsymbol{\bar{\epsilon}}_t-\boldsymbol{\epsilon}_\theta(\mathbf{x}_t, t)\right\|^2,
\end{equation}
the term $\boldsymbol{\epsilon}_\theta$ represents the trainable neural network, and this training scheme is referred to as the $\text{pred}-\epsilon$ mode.

When the mean $\tilde{\mu}_t$ of the posterior distribution is expressed in the form of Eq.(\ref{deqn_ex7a}), the only unknown variable is $\mathbf{x}_0$. Therefore, a neural network is employed to predict $\mathbf{x}_0$, and the corresponding loss function is formulated as follows:
\begin{equation}
\label{deqn_ex10a}
L=\left\|\mathbf{x}_0-\mathbf{x}_\theta(\mathbf{x}_t, t)\right\|^2,
\end{equation}
the term $\mathbf{x}_\theta$ represents the trainable neural network, and this training scheme is referred to as the $\text{pred}-\mathbf{x}_0$ mode.

The generation process of the diffusion model corresponds to the reverse denoising procedure, in which a sample $\mathbf{x}_T$ is first drawn from a standard Gaussian distribution $\mathcal{N}(\mathbf{0}, \mathbf{I})$. Thereafter, the model employs the learned approximate posterior $p_\theta(\mathbf{x}_{t-1}\!\mid\!\mathbf{x}_{t})$ to iteratively generate samples through a Markovian reverse process. The single-step sampling formulations for the $\text{pred}-\epsilon$ and $\text{pred}-\mathbf{x}_0$ are given in Eq.(\ref{deqn_ex11a}) and Eq.(\ref{deqn_ex12a}), respectively.
\begin{equation}
\label{deqn_ex11a}
\mathbf{x}_{t-1}=\frac{1}{\sqrt{\alpha_t}}\left(\mathbf{x}_t-\frac{1-\alpha_t}{\sqrt{1-\bar{\alpha}_t}} \boldsymbol{\epsilon}_\theta(\mathbf{x}_t, t)\right)+\tilde{\beta}_t \boldsymbol{\epsilon},
\end{equation}
\begin{equation}
\label{deqn_ex12a}
\mathbf{x}_{t-1}=\frac{\sqrt{\alpha_t}(1-\bar{\alpha}_{t-1}) \mathbf{x}_t+\sqrt{\bar{\alpha}_{t-1}}(1-\alpha_t) \mathbf{x}_\theta(\mathbf{x}_t, t)}{1-\bar{\alpha}_t}+\tilde{\beta}_t \boldsymbol{\epsilon}.
\end{equation}
After $T$ reverse steps, the final generated sample can be obtained. Empirically, it has been observed that the conditional diffusion model for CGM generation achieves superior performance under the $\text{pred}-\mathbf{x}_0$ mode compared with the $\text{pred}-\epsilon$ mode. Therefore, both the training and sampling procedures in this paper adopt the $\text{pred}-\mathbf{x}_0$ mode.

\subsection{Denoising Diffusion Implicit Model}
The DDPM requires $T$ iterative steps to generate a single sample, which results in relatively slow generation speed. To address this limitation, the Denoising Diffusion Implicit Model (DDIM)\cite{Song2020DenoisingDI} removes the constraint that the reverse denoising process must follow a Markov chain, thereby achieving significantly faster sampling while maintaining comparable generation quality.

For the posterior distribution in Eq.(\ref{deqn_ex6a}), if the term $q(\mathbf{x}_t \!\mid\! \mathbf{x}_{t-1}, \mathbf{x}_0)$ associated with the Markov process is not invoked, the posterior can be derived via the method of undetermined coefficients as follows:
\begin{equation}
\begin{split}
\label{deqn_ex13a}
q(\mathbf{x}_{t-1} \mid \mathbf{x}_t, \mathbf{x}_0)=\mathcal{N}(\mathbf{x}_{t-1} ; \frac{\sqrt{\bar{\beta}_{t-1}-\sigma_t^2}}{\sqrt{\bar{\beta}_t}} \mathbf{x}_t
+\left(\sqrt{\bar{\alpha}_{t-1}}-\frac{\sqrt{\bar{\alpha}_t} \sqrt{\bar{\beta}_{t-1}-\sigma_t^2}}{\sqrt{\bar{\beta}_t}}\right) \mathbf{x}_0, \sigma_t^2 \mathbf{I}).
\end{split}
\end{equation}
This distribution is referred to as the posterior probability distribution of the generalized diffusion model. By substituting Eq.(\ref{deqn_ex5a}) into Eq.(\ref{deqn_ex13a}), we obtain:
\begin{equation}
\begin{split}
\label{deqn_ex14a}
q(\mathbf{x}_{t-1} \mid \mathbf{x}_t)=\mathcal{N}(\mathbf{x}_{t-1} ; \frac{1}{\sqrt{\alpha_t}}(\mathbf{x}_t-\sqrt{1-\bar{\alpha}_t}
-\sqrt{\alpha_t} \sqrt{1-\bar{\alpha}_t-\sigma_t^2}) \boldsymbol{\epsilon}_t, \sigma_t^2 \mathbf{I}).
\end{split}
\end{equation}
Since the above sampling process does not need to follow a Markov chain, $\mathbf{x}_{t-1}$ can, in fact, correspond to the data $\mathbf{x}_{pre}$ at any previous time step. Therefore, the generalized diffusion model defines the sampling formulation as follows:
\begin{equation}
\begin{split}
\label{deqn_ex15a}
\mathbf{x}_{pre}=\sqrt{\bar{\alpha}_{pre}}\left(\frac{\mathbf{x}_t-\sqrt{1-\bar{\alpha}_t} \boldsymbol{\epsilon}_t}{\sqrt{\bar{\alpha}_t}}\right)
+\sqrt{1-\bar{\alpha}_{pre}-\sigma_t^2} \boldsymbol{\epsilon}_t+\sigma_t^2 \boldsymbol{\epsilon}.
\end{split}
\end{equation}
When $\sigma_t=0$, the process reduces to DDIM; thus, the sampling formulation of DDIM can be expressed as:
\begin{equation}
\label{deqn_ex16a}
\mathbf{x}_{pre}=\sqrt{\bar{\alpha}_{pre}}\left(\frac{\mathbf{x}_t-\sqrt{1-\bar{\alpha}_t} \boldsymbol{\epsilon}_t}{\sqrt{\bar{\alpha}_t}}\right)+\sqrt{1-\bar{\alpha}_{pre}} \boldsymbol{\epsilon}_t.
\end{equation}
In this case, the sampled data $\mathbf{x}_{pre}$ can be separated from $\mathbf{x}_t$ by multiple time steps, thereby accelerating the generation process.
\subsection{Latent Diffusion Model}
The diffusion model commonly employs a UNet architecture as the neural network for predicting either $\boldsymbol{\bar{\epsilon}}_t$ or $\mathbf{x}_0$. However, when the dimensionality of the input data is excessively high, UNet inevitably becomes large in scale, leading to increased computational cost during both training and inference. The Latent Diffusion Model (LDM)\cite{Rombach2021HighResolutionIS} first employs the encoder $\mathcal{E}$ of a Variational Autoencoder (VAE) to compress the high-dimensional input data $\mathbf{x}$ into a low-dimensional latent feature space $\mathbf{z}$. The UNet of the diffusion model is then trained in this latent space, which effectively reduces the network scale and computational cost. During the generation phase, the diffusion model produces a latent representation $\mathbf{z}_0$, which is subsequently decoded by the VAE decoder $\mathcal{D}$ to reconstruct the desired high-dimensional data $\mathbf{x}_0$. 

In addition, the LDM introduces a novel conditional guidance mechanism that incorporates conditional prompts through a cross-attention module. This mechanism not only supports multimodal conditional inputs but also enables more precise conditional control, thereby providing a more flexible and effective approach to conditional guidance. The external conditional prompt $y$ is first encoded by the conditional encoder $\tau_\theta$ to obtain $\tau_\theta(y) \in \mathbb{R}^{n \times d_i}$. The internal feature tensor $\mathbf{z}_t$ is flattened to obtain the $\varphi(\mathbf{z}_t) \in \mathbb{R}^{n \times d_c}$. They are then multiplied by the matrices $W_Q \in \mathbb{R}^{d_c \times d_n}$, $W_K \in \mathbb{R}^{d_i \times d_n}$, $W_V \in \mathbb{R}^{d_i \times d_n}$ to obtain $Q$, $K$, and $V$, respectively, as follows:
\begin{equation}
\label{deqn_ex17a}
Q=\varphi(\mathbf{z}_t) \cdot W_Q, K=\tau_\theta(y) \cdot W_K, V=\tau_\theta(y) \cdot W_V.
\end{equation}
The computation of the cross-attention mechanism is formulated as follows:
\begin{equation}
\label{deqn_ex18a}
\text{CrossAtten}=\operatorname{softmax}\left(\frac{Q \cdot K^T}{\sqrt{d_n}}\right) \cdot V.
\end{equation}
During the second-stage training of the LDM, the conditional encoder is jointly optimized with the UNet. Therefore, the loss function for the second-stage training of the LDM is defined as follows:
\begin{equation}
\label{deqn_ex19a}
L_{\text{LDM}}=\left\|\bar{\boldsymbol{\epsilon}}_t-\boldsymbol{\epsilon}_\theta\left(\mathbf{z}_t, t, \tau_\theta(y)\right)\right\|^2.
\end{equation}

\section{Physics-inspired Channel Knowledge Map Generation Diffusion Model}
\subsection{The Physical Meaning and Calculation Method of the Proposed Physical Regularization Terms}
To incorporate CGM-specific physical constraints into the learning process, we design three novel physics-inspired regularization terms, which are described as follows.
\subsubsection{Edge Loss Term}
\begin{figure}[!htbp]
\centering
\includegraphics[width=2.5in]{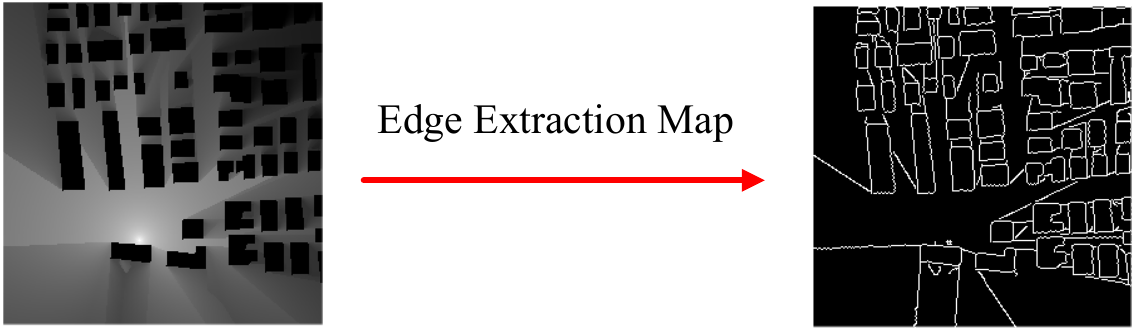}
\caption{Edge Extraction Illustration.}
\label{fig_2}
\end{figure}

By observing the heatmap of the CGM, it can be found that its spatial distribution is highly consistent with an image captured under point-light illumination. Distinct edge structures appear in regions where obstacles cause signal occlusion. Therefore, as illustrated in Fig. \ref{fig_2}, edge extraction can be applied to the CGM, and the resulting edge map depicts partial propagation paths of wireless signals after interacting with obstacles such as buildings. 

The residual between the edge map of the ground truth CGM and that of the generated CGM is defined as the edge loss. This loss term ensures that the propagation paths from the BS to various spatial locations in the generated CGM are consistent with the surrounding environment and conform to the physical constraint that large-scale wireless signal attenuation is primarily dominated by the LoS component.There exist various methods for edge detection. Considering computational efficiency, we adopt the Laplace operator to perform edge extraction on the CGM. The discrete form of the Laplace operator used for edge extraction is expressed as follows:
\begin{equation}
\begin{split}
\label{deqn_ex20a}
\nabla^2 \mathbf{x}(i, j)=\frac{\mathbf{x}(i+1, j)+\mathbf{x}(i-1, j)+\mathbf{x}(i, j+1)}{h^2}
+\frac{\mathbf{x}(i, j-1)-4\mathbf{x}(i, j)}{h^2},
\end{split}
\end{equation}
where $h$ represents the grid interval. After applying the Laplace operator to the CGM, a binary edge map can be obtained by thresholding the result. However, in deep learning, the binarization operation blocks gradient backpropagation, preventing the model from being trained end-to-end. To address this issue, we employ an approximate gradient propagation technique known as the Straight-Through Estimator (STE)\cite{bengio2013estimating} to compute the edge loss term. When the input is denoted as $\mathbf{x}$, the STE function is defined as follows:
\begin{equation}
\label{deqn_ex21a}
\begin{aligned}
& \mathbf{soft}=\operatorname{sigmoid}\left(\alpha\left(\frac{\mathbf{x}}{\max(\mathbf{x})}-\tau\right)\right) \\
& \mathbf{hard}=\left(\mathbf{soft}>0.5\right) \\
& \operatorname{STE}\left(\mathbf{x}\right)=\mathbf{hard}+\mathbf{soft}-\mathbf{soft}\operatorname{.detach()},
\end{aligned}
\end{equation}
here, $\alpha$ is a parameter that controls the steepness of the Sigmoid function, and $\tau$ denotes the threshold. The operation $\operatorname{.detach()}$ indicates that the gradient propagation of the corresponding tensor is stopped. Consequently, $\operatorname{STE(x)}$ behaves as a $\mathbf{hard}$ (binarized) function during the forward pass, whereas it adopts the $\mathbf{soft}$ gradient during backpropagation. The final formulation of the edge loss term is expressed as follows:
\begin{equation}
\label{deqn_ex22a}
L_{\text{edge}}=\frac{1}{N} \sum_{n=1}^N\left\|\operatorname{STE}(\nabla^2 \mathbf{x}_{0,n})-\operatorname{STE}(\nabla^2 \hat{\mathbf{x}}_{0,n})\right\|^2,
\end{equation}
where $\mathbf{x}_{0,n}$ denotes the ground truth CGM samples, $\hat{\mathbf{x}}_{0,n}$ represents the CGM samples generated during training, and $N$ indicates the total number of samples.
\subsubsection{Regional Propagation Loss Term}
If there are no obstacles within a region, the path loss increases logarithmically with propagation distance. However, in the presence of obstacles, the spatial distribution of the CGM becomes nonuniform and strongly correlated with the locations of these obstacles. By examining the heatmap of the CGM, it can be observed that the map is segmented into multiple color blocks due to the influence of buildings and other obstructions. Areas with no obstacles between the BS and the receiver appear as lighter color regions, indicating smaller path losses, whereas regions with multiple obstacles appear darker, representing larger path losses.

\begin{figure}[!htbp]
\centering
\includegraphics[width=2.5in]{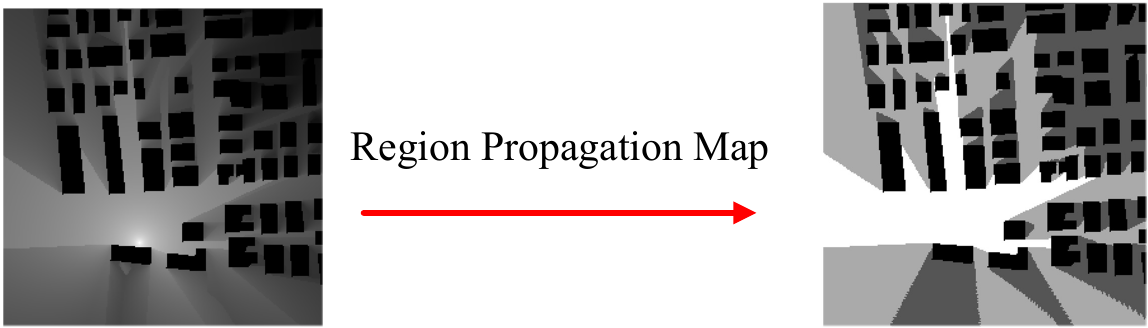}
\caption{Regional Propagation Illustration.}
\label{fig_3}
\end{figure}

Inspired by this observation, we divide the CGM into four regions according to the level of path loss: LoS region, slightly shadowed region, heavily shadowed region, and building region. The propagation regions are determined based on the difference between the path loss observed in the CGM and that computed using the logarithmic distance path-loss model. A larger difference indicates stronger obstruction, while almost no difference implies an unobstructed LoS area. The resulting regional propagation map is shown in Fig. \ref{fig_3}.

\begin{figure}[!htbp]
\centering
\includegraphics[width=3.4in]{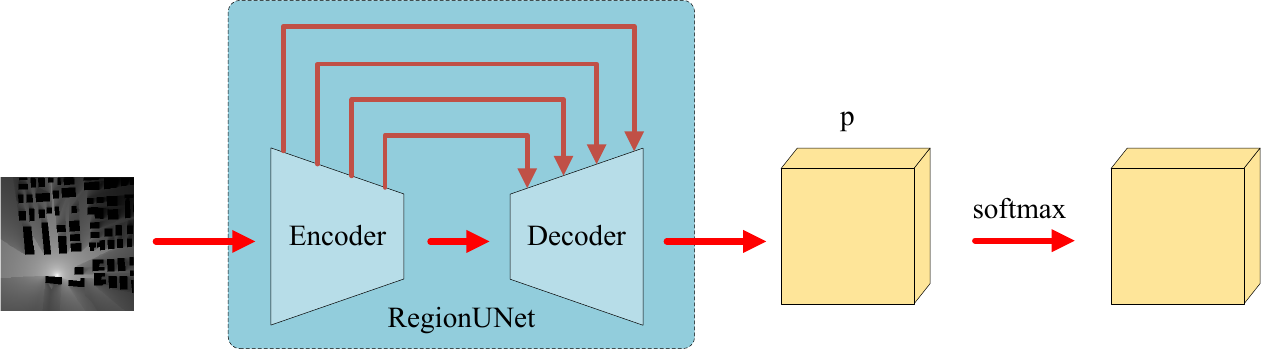}
\caption{Computation of Regional Propagation Map Using RegionUNet.}
\label{fig_4}
\end{figure}

In this work, a neural network is employed to directly map the CGM to its corresponding regional propagation map. As described earlier, the process of regional propagation division is analogous to image segmentation in computer vision. Therefore, we adopt the UNet architecture—renowned for its strong segmentation capability—to perform this task, referred to here as RegionUNet. During training, the target regional maps in the dataset are pre-obtained using the aforementioned multi-threshold division method. After training, the network parameters of RegionUNet are fixed, and the network is then used during the diffusion model training stage to compute the regional propagation maps of both the ground truth and generated CGMs, as illustrated in Fig. \ref{fig_4}.

To facilitate the computation of the regional propagation loss term, the output tensor of RegionUNet is denoted as $p \in R^{4 \times H \times H}$. A softmax operation is applied along the channel dimension of $p$, so that each element of the resulting tensor represents the probability of a given position belonging to a specific propagation region. Accordingly, the regional propagation loss term is formulated as follows:
\begin{equation}
\begin{split}
\label{deqn_ex23a}
L_{\text{region}}=\frac{1}{N} \sum_{n=1}^N\left\|\operatorname{R}(\mathbf{x}_{0, n})-\operatorname{R}(\hat{\mathbf{x}}_{0, n})\right\|^2,
\end{split}
\end{equation}
where $\operatorname{R}(\cdot)=\operatorname{softmax}(\operatorname{UNet}(\cdot))$.
\subsubsection{Multi-scale Feature Loss Term}
The CGM can be regarded as a spatial field in which each point represents the signal strength at a specific location. The Laplacian pyramid enables the decomposition of this spatial field into multiple layers, each capturing signal variation patterns at a distinct spatial scale. Specifically, each layer of the Laplacian pyramid corresponds to the variation information of the CGM at a particular spatial scale.

\begin{figure}[!htbp]
\centering
\includegraphics[width=3.4in]{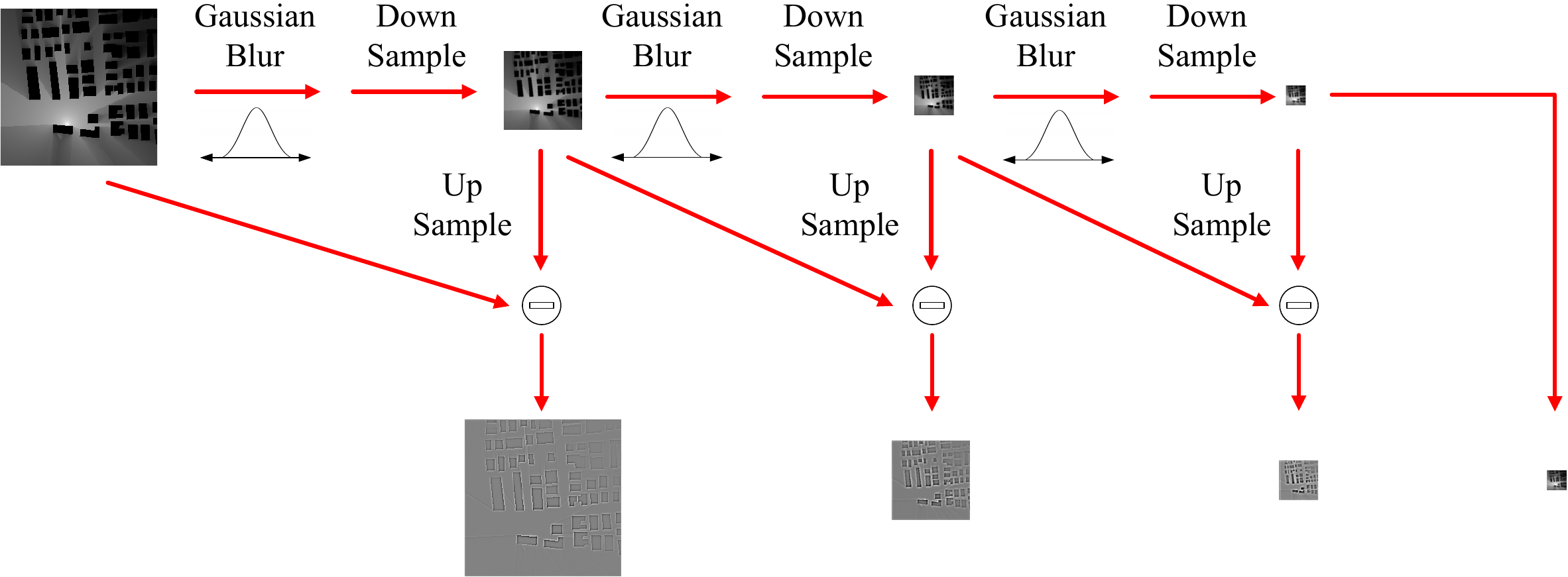}
\caption{Multi-Scale Feature Map Computation Based on the Laplacian Pyramid.}
\label{fig_5}
\end{figure}

The lower layers capture fine-scale variations, such as the rapid signal attenuation at building edges or narrow streets caused by local shadowing and diffraction effects. The middle layers reflect medium-scale variations, representing path loss changes across different urban blocks and propagation behaviors within urban canyons. The upper layers characterize large-scale trends, describing the overall propagation tendency of the signal, i.e., large-scale fading in free-space environments. Therefore, by applying the Laplacian pyramid to both the ground truth and generated CGMs, we extract their multi-scale feature maps, and the residual between these feature maps is incorporated into the training process as a physics-inspired regularization term. This design ensures that the generated CGM conforms to the multi-scale physical effects inherent in wireless signal propagation. The procedure for multi-scale feature extraction using the Laplacian pyramid is illustrated in Fig. \ref{fig_5}.

The input to the first layer of the Laplacian pyramid is the CGM. At layer $l$, the input is denoted by $\mathbf{x}_l$, the output $\mathbf{x}_{l+1}$ is obtained by applying Gaussian smoothing to $\mathbf{x}_l$ followed by $2\times$ downsampling. The feature map at scale $l$ is then computed as the difference between $\mathbf{x}_l$ and the $2\times$ upsampled version of $\mathbf{x}_{l+1}$. Therefore, the multi-scale feature loss term is computed by performing Laplacian pyramid decomposition, as shown in Fig. \ref{fig_5}, on both the ground truth and generated CGMs. The residuals between the corresponding feature maps at each scale are then summed to obtain the overall loss. The calculation formula is expressed as follows:
\begin{equation}
\label{deqn_ex24a}
L_{\text{mul-fea}}=\sum_{l=1}^L w_l \sum_{n=1}^N\left\|\mathcal{L}_l(\mathbf{x}_{0, n})-\mathcal{L}_l(\hat{\mathbf{x}}_{0, n})\right\|^2,
\end{equation}
where, $\mathcal{L}_l(\mathbf{x})$ denotes the feature map extracted from the l-th layer of the Laplacian pyramid, and $w_l$ represents the loss weight assigned to the feature map at the l-th scale.
\subsection{The Derivation of Model Loss Function}
Discriminative models, such as autoencoders, UNet, and LSTM, establish a direct mapping from input to output; therefore, their loss functions are typically defined as the Mean Squared Error (MSE) between the input and the output. When constructing a physics-inspired model based on a discriminative framework, the physics-inspired terms can be directly appended to the original MSE loss function, since both components jointly optimize the mapping relationship from input to output. However, as described in Subsection III A, the neural network $p_\theta(\mathbf{x}_{t-1}\!\mid\!\mathbf{x}_t)$ in diffusion models is designed to approximate the true posterior distribution $q(\mathbf{x}_{t-1}\!\mid\!\mathbf{x}_t, \mathbf{x}_0)$. Although, under the $\text{pred}-\mathbf{x}_0$ mode, the loss function can be simplified to the MSE between the network output and the input sample in form, the neural network in this mode still aims to fit the mean of the posterior distribution. Therefore, its output cannot be directly used as the generated sample. Since the optimization objective of the loss function in the vanilla diffusion model does not align with that required by our designed physics-inspired regularization terms, the loss function formulation of the physics-guided diffusion model needs to be rederived.

We define the physical residual of the model as:
\begin{equation}
\label{deqn_ex25a}
r(\hat{\mathbf{x}}_0)=\left[\begin{array}{c}
\operatorname{STE}(\nabla^2 \mathbf{x}_0)-\operatorname{STE}(\nabla^2 \hat{\mathbf{x}}_0) \\
\operatorname{R}(\mathbf{x}_0)-\operatorname{R}(\hat{\mathbf{x}}_0) \\
\mathcal{L}_{l}(\mathbf{x}_0)-\mathcal{L}_{l}(\hat{\mathbf{x}}_0) \\
\end{array}\right].
\end{equation}

Then, we define a virtual observation $\tilde{r}$ to introduce uncertainty into the model’s physical residual $r(\hat{\mathbf{x}}_0)$\cite{RIXNER2021110218}. The distribution of $\tilde{r}$ is given as follows:
\begin{equation}
\label{deqn_ex26a}
q_R(\tilde{r} \mid \hat{\mathbf{x}}_0)=\mathcal{N}\left(\tilde{r} \mid r(\hat{\mathbf{x}}_0), \sigma^2 \mathbf{I}\right),
\end{equation}
here, $\sigma$ represents the constraint strength: a smaller value indicates a stronger constraint imposed by the physical model, while a larger value implies a weaker constraint. The virtual observation $\tilde{r}$ enables the physical residual to be incorporated into the generative model within a variational framework, allowing the physical law to act as a probabilistic observation during training. Accordingly, the loss function of the physics-inspired term during training can be expressed as follows:
\begin{equation}
\label{deqn_ex27a}
\arg \max _\theta E_{\mathbf{x}_{1:T}\sim q(\mathbf{x}_{1:T})}\left[\log q_R\left(\tilde{r}=0 \mid \hat{\mathbf{x}}_0(\mathbf{x}_{1:T})\right)\right].
\end{equation}

Considering that the loss function of the diffusion model aims to maximize the cross-entropy between the true data distribution and the modeled data distribution, the overall loss function of the proposed physics-inspired diffusion model can be expressed as follows:
\begin{equation}
\label{deqn_ex28a}
\begin{aligned}
&\arg \max_\theta E_{\mathbf{x}_0 \sim q(\mathbf{x}_0)}\left[\log p_\theta(\mathbf{x}_0)\right]
+ E_{\mathbf{x}_{1:T} \sim q(\mathbf{x}_{1:T})}\left[\log q_R(\tilde{r}=0 \mid \hat{\mathbf{x}}_0(\mathbf{x}_{1:T}))\right] \\
= &\arg \min_\theta E_{\mathbf{x}_0 \sim q(\mathbf{x}_0)}\left[-\log p_\theta(\mathbf{x}_0)\right]
+ E_{\mathbf{x}_{1:T} \sim q(\mathbf{x}_{1:T})}\left[-\log q_R(\tilde{r}=0 \mid \hat{\mathbf{x}}_0(\mathbf{x}_{1:T}))\right].
\end{aligned}
\end{equation}

Based on the construction of the variational lower bound in\cite{ho2020denoising} for deriving the DDPM loss function, the numerical formulation of the loss function for the proposed physics-inspired diffusion model can be derived as follows:
\begin{equation}
\label{deqn_ex29a}
\begin{aligned}
    & E_{\mathbf{x}_0 \sim q\left(\mathbf{x}_0\right)}\left[-\log p_\theta\left(\mathbf{x}_0\right)\right]
    +E_{\mathbf{x}_{1:T} \sim q\left(\mathbf{x}_{1:T}\right)}\left[-\log q_R\left(\tilde{r}=0 \mid \hat{\mathbf{x}}_0\left(\mathbf{x}_{1:T}\right)\right)\right] \\
    \leq &E_{q\left(\mathbf{x}_{0:T}\right)}\left[\log \frac{q\left(\mathbf{x}_{1:T} \mid \mathbf{x}_0\right)}{p_\theta\left(\mathbf{x}_{0:T}\right)}\right]
    +E_{\mathbf{x}_{1:T} \sim q\left(\mathbf{x}_{1:T}\right)}\left[-\log q_R\left(\tilde{r}=0 \mid \hat{\mathbf{x}}_0\left(x_{1:T}\right)\right)\right] \\
    = &E_{q\left(\mathbf{x}_{0:T}\right)}\left[\log \frac{q\left(\mathbf{x}_{1: T} \mid \mathbf{x}_0\right)}{p_\theta\left(\mathbf{x}_{0: T}\right)}-\log q_R\left(\tilde{r}=0 \mid \hat{\mathbf{x}}_0\left(\mathbf{x}_{1:T}\right)\right)\right] \\
    = &E_{t\sim[1,T],\mathbf{x}_{0:T}\sim q\left(\mathbf{x}_{0:T}\right)}[\frac{\tilde{A}_t^2}{2 \tilde{\beta}_t^2}\left\|\mathbf{x}_0-\mathbf{x}_\theta\left(\mathbf{x}_t, t\right)\right\|^2
    +\frac{1}{2 \sigma^2}\left\|r\left(\hat{\mathbf{x}}_0\left(\mathbf{x}_t, t\right)\right)\right\|^2].
\end{aligned}
\end{equation}
The detailed derivation of Eq.(\ref{deqn_ex29a}) is provided in the Appendix, where $\tilde{A}_t=\frac{\sqrt{\bar{\alpha}_{t-1}} \beta_t}{1-\bar{\alpha}_t}$, $\tilde{\beta}_t=\frac{1-\bar{\alpha}_{t-1}}{1-\bar{\alpha}_t} \beta_t$. Similar to\cite{ho2020denoising}, we observe that ignoring the preceding weights leads to better generative performance. Therefore, the final loss function is formulated as follows:
\begin{equation}
\label{deqn_ex30a}
\begin{aligned}
L_{\text{phy}}(\theta)=E_{t\sim[1,T],\mathbf{x}_{0:T} \sim q\left(\mathbf{x}_{0:T}\right)}[\left\|\mathbf{x}_0-\mathbf{x}_\theta\left(\mathbf{x}_t, t\right)\right\|^2
+\left\|r\left(\hat{\mathbf{x}}_0\left(\mathbf{x}_t, t\right)\right)\right\|^2].
\end{aligned}
\end{equation}

\begin{algorithm}[H]
\caption{DDIM Sampling algorithm}\label{alg:alg1}
\begin{algorithmic}[1]
\Require Select a subsequence of length $I+1$ from the sequence $\{1,2, \ldots, T\}$: $\left\{t_{s_0}=1, t_{s_1}, \ldots, t_{s_{I-1}}, t_{s_I}=T\right\}$, input the condition UNet $\mathbf{z}_\theta$, the VAE decoder $\mathcal{D}$ and the conditional prompt $\mathbf{c}$
\State $\mathbf{z}_T \sim \mathcal{N}(\mathbf{0}, \mathbf{I})$
\For{$i$ in $\{I, I-1, I-2, \ldots, 2,1,0\}$}
    \If{$i==0$}
        \State $\hat{\mathbf{z}}_0=\mathbf{z}_\theta\left(\mathbf{z}_1, 1, \mathbf{c}\right)$
    \Else
        \State $\tilde{\mathbf{\epsilon}}_\theta=\frac{\mathbf{z}_{t_{s_i}}}{\sqrt{1-\bar{\alpha}_{t_{s_i}}}}-\frac{\sqrt{\bar{\alpha}_{t_{s_i}}}}{\sqrt{1-\bar{\alpha}_{t_{s_i}}}} \mathbf{z}_\theta\left(\mathbf{z}_{t_{s_i}}, t_{s_i}, \mathbf{c}\right)$
        \State $\mathbf{z}_{t_{s_{i-1}}}=\sqrt{\bar{\alpha}_{t_{s_{i-1}}}} \mathbf{z}_\theta\left(\mathbf{z}_{t_{s_i}}, t_{s_i}, \mathbf{c}\right)+\sqrt{1-\bar{\alpha}_{t_{s_{i-1}}}} \tilde{\mathbf{\epsilon}}_\theta$
    \EndIf
\EndFor
\State $\hat{\mathbf{x}}_0=\mathcal{D}(\hat{\mathbf{z}}_0)$
\State \Return $\hat{\mathbf{x}}_0$
\end{algorithmic}
\end{algorithm}

\subsection{The Physics-inspired Channel Knowledge Map Generation Diffusion Model Architecture and Algorithm Flow}
\begin{figure*}[htbp]
  \centering
  \subfloat[]{
      \centering
      \includegraphics[width=3in]{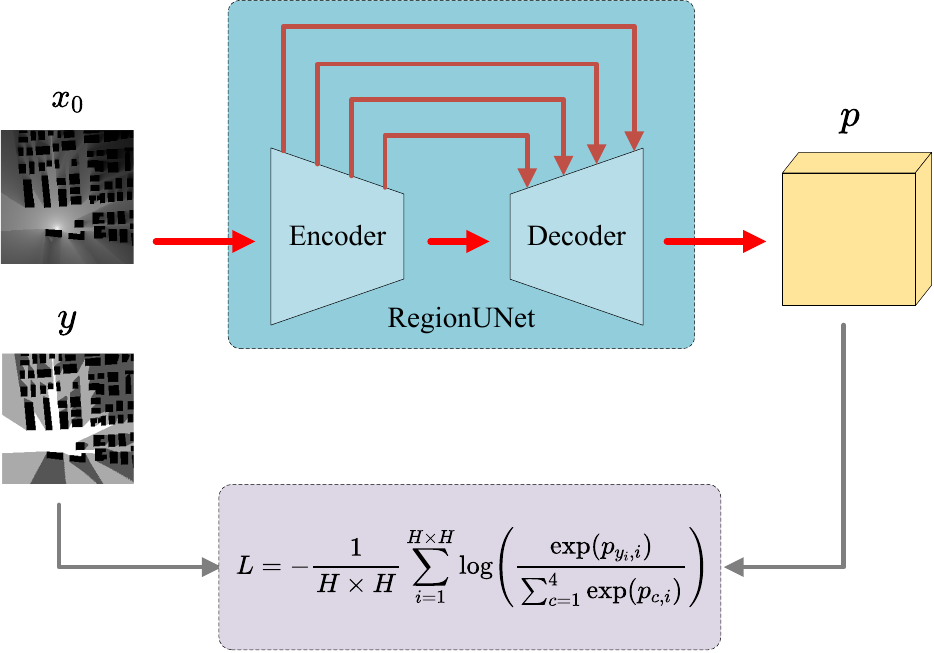}
      \label{fig_6a}}
  \hfil
  \subfloat[]{
      \centering
      \includegraphics[width=3in]{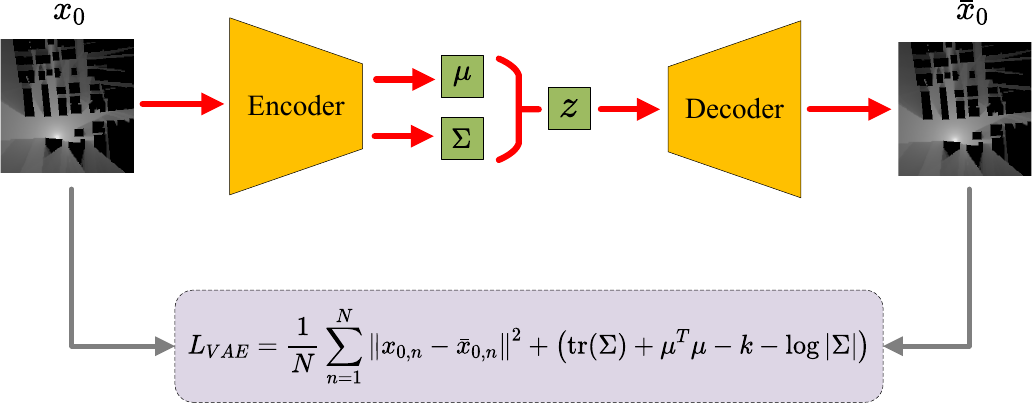}
      \label{fig_6b}}

  \subfloat[]{
      \centering
      \includegraphics[width=6in]{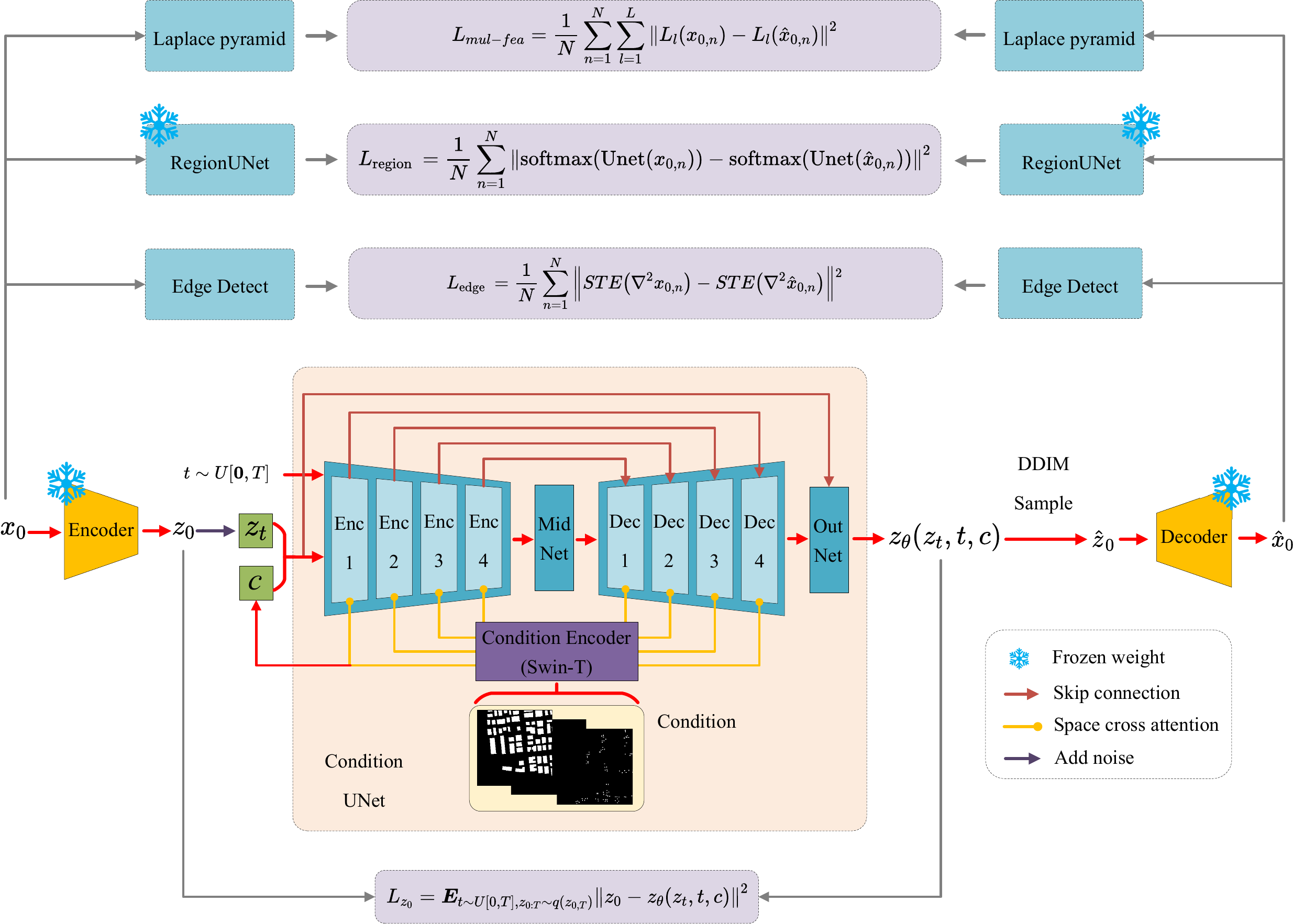}
      \label{fig_6c}}
  
  \caption{Training Architecture of the Physics-inspired Channel Knowledge Map Generation Diffusion Model. (a) Stage 1: Training of RegionUNet. (b) Stage 2: Training of VAE. (c) Stage 3: Training of the Diffusion Model.}
  \label{fig_6}
\end{figure*}

The overall training framework of the proposed physics-inspired generation diffusion model is illustrated in Fig. (\ref{fig_6}). As shown in Fig. (\ref{fig_6a}), the first stage trains the RegionUNet network, which is employed to compute the regional propagation loss term. The second stage involves training a VAE that compresses the input data into a low-dimensional latent tensor, and the output of the VAE is defined as $\bar{\mathbf{x}}_0$, as depicted in Fig. (\ref{fig_6b}). After the completion of the first two stages, the network parameters of both the RegionUNet and the VAE are frozen and excluded from the training process in the third stage. The training framework of the diffusion model in the third stage is illustrated in Fig. (\ref{fig_6c}). The conditional prompt $\mathbf{c}$ includes the building layout map, the BS location map, and the vehicle location map. For the conditional encoder, we employ the Swin-T (Shifted Window Transformer) network, which has demonstrated strong performance in image processing tasks. The conditional prompts are encoded by the Swin-T and injected into the main UNet backbone through a cross-attention mechanism. The UNet backbone and the Swin-T encoder are jointly trained. During training, in addition to the original loss term $L_{z_0}$ defined in the LDM, three physics-inspired regularization terms, $L_{\text{edge}}$, $L_{\text{region}}$, $L_{\text{mul-fea}}$, are also incorporated into the total loss function. Therefore, unlike other diffusion models, each computation of the total loss function $L_{\text{total}}$ requires performing a reverse denoising process to generate the CGM. However, employing the reverse denoising sampling procedure of the DDPM to obtain the generated CGM would result in an unacceptably long training time. To address this issue, we adopt the DDIM during training to generate CGM $\hat{\mathbf{x}}_0$. The algorithmic procedure for DDIM sampling under the $\text{pred}-\mathbf{x}_0$ mode is presented in Algorithm \ref{alg:alg1}. 

The training algorithm of the proposed model is summarized in Algorithm \ref{alg:alg2}. In line 13 of Algorithm \ref{alg:alg2}, the three physics-based regularization terms are each multiplied by their respective weighting coefficients. Through empirical experiments, we observed that the model achieves the best training performance when the weighting coefficients are set to $\lambda_{\text{edge}}=10^{-3}$, $\lambda_{\text{region}}=5\times10^{-3}$, and $\lambda_{\text{mul-fea}}=5\times10^{-3}$.

\begin{algorithm}[H]
\caption{Training algorithm}\label{alg:alg2}
\begin{algorithmic}[1]
\Require Input $\mathbf{x}_0, \mathbf{c} \sim Dataset$, the frozen VAE encoder $\mathcal{E}$, the frozen RegionUNet
\Repeat
    \State $\mathbf{x}_0 \sim q\left(\mathbf{x}_0\right)$
    \State $\mathbf{z}_0=\mathcal{E}(\mathbf{x}_0)$
    \State $t \sim \operatorname{Uniform}\{1, \ldots, T\}$
    \State $\mathbf{\epsilon} \sim \mathcal{N}(\mathbf{0}, \boldsymbol{I}) $
    \State $\mathbf{z}_t=\sqrt{\bar{\alpha}_t} \mathbf{z}_0+\sqrt{1-\bar{\alpha}_t} \mathbf{\epsilon} $
    \State Obtain $\hat{\mathbf{x}}_0$ using DDIM Sampler
    \State $L_{z_0}=\left\|\mathbf{z}_0-\mathbf{z}_\theta(\mathbf{z}_t, t, \mathbf{c})\right\|^2$
    \State $L_{\text{edge}}=\left\|\operatorname{STE}\left(\nabla^2 \mathbf{x}_{0}\right)-\operatorname{STE}\left(\nabla^2 \hat{\mathbf{x}}_{0}\right)\right\|^2$
    \State $L_{\text{region}}=\left\|\operatorname{R}\left(\mathbf{x}_{0}\right)-\operatorname{R}\left(\hat{\mathbf{x}}_{0}\right)\right\|^2$
    \State where $\operatorname{R}(\cdot)=\operatorname{softmax}\left(\operatorname{UNet}\left(\cdot\right)\right)$
    \State $L_{\text{mul-fea}}=\left\|\mathcal{L}_l\left(\mathbf{x}_{0}\right)-\mathcal{L}_l\left(\hat{\mathbf{x}}_{0}\right)\right\|^2$
    \State $L_{\text{total}}=L_{z_0}+\lambda_{\text{edge}}L_{\text{edge}}+\lambda_{\text{region}}L_{\text{region}}$
    \State \qquad\quad $+\lambda_{\text{mul-fea}}L_{\text{mul-fea}}$
    \State Take gradient descent step on $\nabla_\theta L_{\text{total}}$
\Until{converged}
\end{algorithmic}
\end{algorithm}

\begin{figure*}[!t]
\centering
\includegraphics[width=6.in]{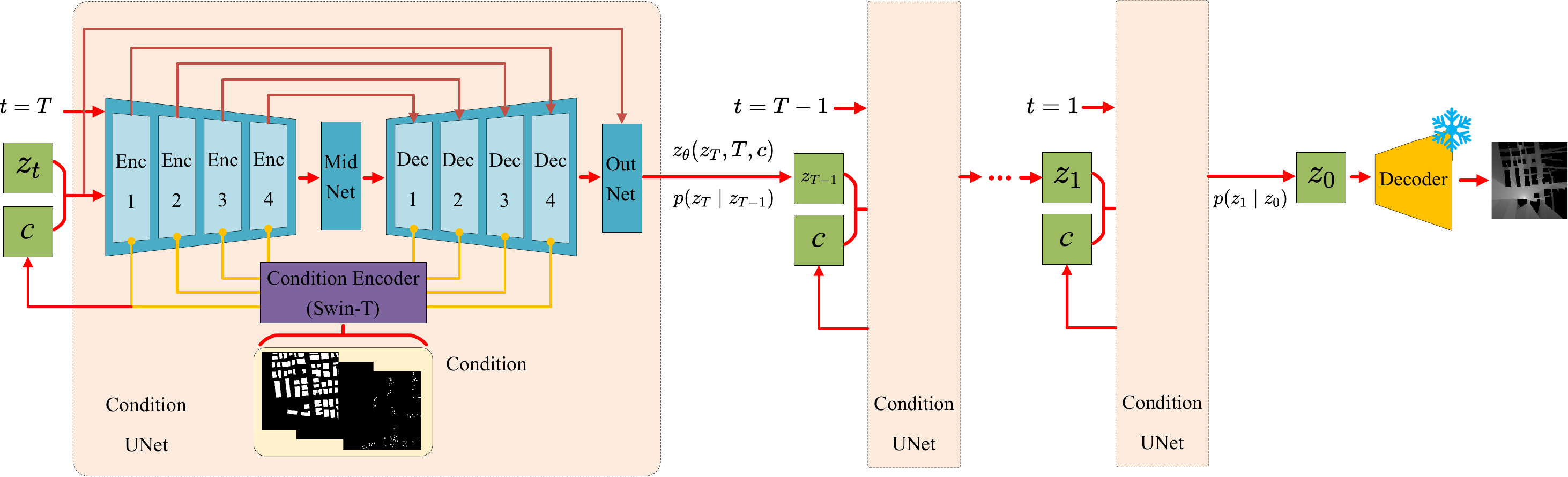}
\caption{Generation Architecture of the Physics-inspired Channel Knowledge Map Generation Diffusion Model.}
\label{fig_7}
\end{figure*}

After completing the three-stage training process, the proposed model is capable of constructing the CGM solely based on the propagation environment. The structural framework of CGM generation using the proposed model is illustrated in Fig. (\ref{fig_7}). We find that, when an appropriate sub-sequence length is selected, the CGMs generated by the DDIM sampling process exhibit nearly identical accuracy to those generated by the DDPM sampling process. Since the DDPM sampling process requires $T$ iterative denoising steps, resulting in excessive generation time, the DDIM sampling algorithm is adopted for CGM generation to meet the speed requirements of practical applications. 

\section{Experiment Results}
\subsection{Experiment Setup}
\subsubsection{Dataset}
In this paper, the open-source dataset named RadioMapSeer\cite{0gtx6v3022} is utilized. The dataset was collected from 700 urban maps in European cities. For each city map, a single BS was deployed, and Winprop was employed to perform ray-tracing channel simulations and generate the corresponding CGM heatmaps. Each CGM is discretized into a $256 \times 256$ grid, where each grid cell represents the path loss value at the corresponding location. RadioMapSeer provides three groups of datasets obtained by different simulation methods, namely DPM (Dominant Path Model), IRT2 (Intelligent Ray Tracing with 2 interactions), and IRT4 (Intelligent Ray Tracing with 4 interactions). For each method, both static CGMs (without vehicles) and dynamic CGMs (with vehicles) were simulated. In the DPM-based dataset, 80 different BS locations were randomly selected for each urban map, resulting in a total of $80 \times 700 = 56000$ CGMs. In our experiments, we use the DPM-generated dataset, which is divided into 40,000 CGMs for training, 8,000 CGMs for validation, and 8,000 CGMs for testing.

\subsubsection{Implentation Details}
In the first training stage, RegionUNet is trained using the Adam optimizer with a batch size of 16 and 100 epochs, and the initial learning rate is set to $1 \times 10^{-4}$. In the second stage, the VAE compresses the input tensor of size (1,256,256) into a latent feature space of dimension (3,64,64). The model is trained using the AdamW optimizer. To reduce GPU memory consumption, gradient accumulation is employed. Specifically, a mini-batch size of 2 is used, and gradients are accumulated over two iterations before performing a backward update, resulting in an effective batch size of 4. The total number of training steps is set to 150,000, which corresponds to approximately 15 effective epochs. The initial learning rate is $1 \times 10^{-5}$. In the third training stage, the number of diffusion timesteps $T$ is set to 1000, and the hyperparameter $\beta_t$ is computed using the cosine scheduling strategy. The optimizer and learning rate settings are consistent with those used for training the VAE. Gradient accumulation is also employed, resulting in an effective batch size of 72 and an equivalent of 180 training epochs. Moreover, our experimental results indicate that the number of sampling steps used by DDIM to generate $\hat{\mathbf{x}}_0$ has a negligible impact on the final training performance. Therefore, in practical training, we typically set the length of the sub-sequence to 2, i.e., $I=1$. The weighting coefficients of each physical regularization term are set as described in Section III, Subsection C, with $\lambda_{\text{edge}}=10^{-3}$, $\lambda_{\text{region}}=5\times10^{-3}$, and $\lambda_{\text{mul-fea}}=5\times10^{-3}$. The training is conducted on a single NVIDIA RTX 5090 GPU and the total training time is approximately 280 hours. During the generation process, by balancing the reconstruction quality of the CGM and the generation time, the DDIM sampler is configured with a subsequence of 5 sampling steps.

\subsubsection{Evaluation Metric}
Since our primary focus is on the reconstruction accuracy of the CGM, three evaluation metrics derived from the MSE are adopted, namely Normalized Mean Squared Error (NMSE), Root Mean Squared Error (RMSE), and Peak Signal-to-Noise Ratio (PSNR).

The NMSE is obtained by normalizing the MSE. In this paper, $\mathbf{x}_n^{pred}$ denotes the reconstructed CGM generated by the model, and $\mathbf{x}_{0,n}$ denotes the ground-truth CGM. The NMSE is calculated as follows:
\begin{equation}
\label{deqn_ex32a}
\text{NMSE}=\frac{1}{N} \sum_{n=1}^N \frac{\text{MSE}_n}{\sum_{i=1}^H \sum_{j=1}^H\left|x_{0, n}(i, j)\right|^2},
\end{equation}
here, the calculation formula of $\text{MSE}_n$ is as follows:
\begin{equation}
\label{deqn_ex33a}
\text{MSE}_n=\sum_{i=1}^H \sum_{j=1}^H\left|x_{0, n}(i, j)-x_n^{\text {pred }}(i, j)\right|^2.
\end{equation}

Since NMSE is normalized, it is a dimensionless metric. The RMSE is the square root of the MSE and therefore has the same physical units as the original data, which makes the error magnitude easier to interpret. Its computation is:
\begin{equation}
\label{deqn_ex34a}
\text{RMSE}=\sqrt{\frac{1}{N} \sum_{n=1}^N \frac{1}{H \times H} \text{MSE}_n}.
\end{equation}

The PSNR expresses the reconstruction error in a logarithmic form from the perspective of the signal-to-noise ratio. It is defined as follows:
\begin{equation}
\label{deqn_ex35a}
\text{PSNR}=\frac{1}{N} \sum_{n=1}^N 10 \lg \left(\frac{\text{MAX}^2}{\text{MSE}_n}\right),
\end{equation}
where MAX denotes the maximum possible value in the CGM. Unlike NMSE and RMSE, a higher PSNR value indicates better reconstruction accuracy.

\subsubsection{Benchmark}
Previous studies have demonstrated that deep learning–based approaches for constructing CGM significantly outperform those based on statistical channel models and spatial interpolation methods. Therefore, in this work, we directly compare our proposed model with three state-of-the-art deep learning–based CGM construction models reported in the literature, namely RadioUNet\cite{9354041}, RME-GAN\cite{10130091}, and RadioDiff\cite{10764739}, to validate the superiority of the proposed method in CGM construction. The details of these three comparison models are described below. 

The RadioUNet model employs two cascaded UNet networks. In\cite{9354041}, RadioUNet has two types of inputs: one that includes sparse measurement values and another that does not, denoted as RadioUNet-s and RadioUNet-c, respectively. Since this paper focuses on CGM construction solely based on the propagation environment, only RadioUNet-c is used for comparison.

The RME-GAN model is based on a GAN framework. Its generator employs both a global loss function and a local loss function. During training, the global loss is first used, followed by the local loss, in order to ensure stable convergence. However, in \cite{10130091}, RME-GAN takes sparse measurement values as part of its input. For a fair comparison in this work, the input to RME-GAN is modified to exclude sparse measurements.

Both RadioDiff and the proposed model are based on the diffusion model framework. However, unlike our model, RadioDiff adopts the training and generation algorithms from the Decoupled Diffusion Model (DDM)\cite{huang2023decoupled}. This framework requires a neural network with a dual UNet architecture. Moreover, to enhance the sharpness of the reconstructed CGM edges, an Adaptive Fast Fourier Transform (FFT) module is introduced between the encoder and decoder of the UNet. As a result, the number of network parameters in RadioDiff is significantly higher than that of the proposed model.

In addition, to ensure a fair comparison, all three baseline models and the proposed model are trained under the same experimental conditions, with identical hyperparameter settings. Therefore, some of the experimental results reported in this paper may differ slightly from those presented in the original publications.

\subsection{Performance Comparison}
\subsubsection{Static Channel Gain Map}
\begin{figure*}[htbp]
\centering
\includegraphics[width=6in]{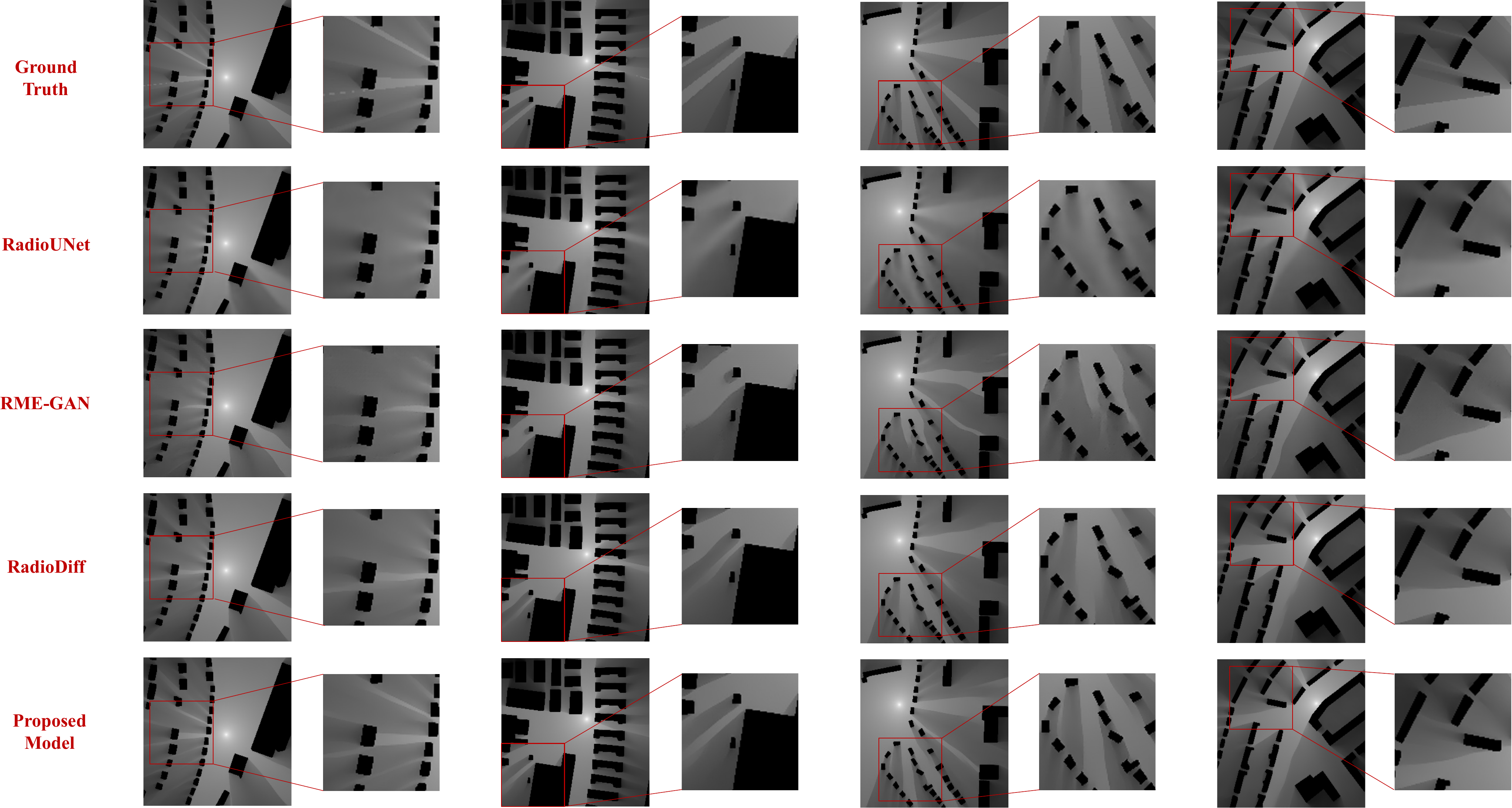}
\caption{Comparison of Static CGMs Constructed by Different Models.}
\label{fig_8}
\end{figure*}

The evaluation metrics of static CGM reconstruction on the test set for different models are presented in Table~\ref{tab1}. It can be observed that the proposed model achieves the best performance across all evaluation metrics. The reconstructed CGMs are visualized in grayscale images, where darker regions indicate higher path loss values. We compare the CGMs reconstructed by different models with the ground-truth CGMs, as shown in Fig. (\ref{fig_8}), where the black regions represent buildings.

\begin{table}[!htbp]
\begin{center}
\caption{Quantitative comparison of static CGM reconstruction results among different models}
\label{tab1}
\begin{tabular}{ m{2cm}<{\centering}  m{1.5cm}<{\centering} m{1.5cm}<{\centering} m{1.5cm}<{\centering}}
\hline
Model & NMSE & RMSE & PSNR\\
\hline
RadioUNet&0.007354&0.024215&32.678948\\
RME-GAN&0.011007&0.030363&30.604957\\ 
RadioDiff&0.005856&0.022150&33.364895\\
Proposed Model&\textbf{0.004697}&\textbf{0.020686}&\textbf{33.940059}\\
\hline
\end{tabular}
\end{center}
\end{table}

As shown in Fig. (\ref{fig_8}), the CGMs reconstructed by RadioUNet, a discriminative model, exhibit very blurry edges and fail to capture the shadowing effect caused by obstacles in wireless signal propagation. The RME-GAN, which is a generative model based on GANs, produces CGMs with sharper edges; however, due to the inherent training instability of GAN and their limited capability in fitting complex data distributions, the reconstructed shadowed regions and edges deviate significantly from the ground-truth CGMs. Consequently, RME-GAN yields the worst performance among all compared models in terms of evaluation metrics. The RadioDiff, leveraging the strong generative capability of diffusion models, alleviates the blurry-edge problem observed in RadioUNet and improves the accuracy of shadowed regions compared with RME-GAN. However, since RadioDiff employs a model originally designed for the image domain, the constructed CGM, though visually plausible, lacks physical realism. In local regions, the edges of occluded areas remain distorted, and the spatial distribution of regions heavily blocked by obstacles is still physically unreasonable. In contrast, the CGMs reconstructed by the proposed physics-inspired model exhibit spatial distributions that are more consistent with the underlying propagation environment. The grayscale details of the reconstructed CGMs better reflect the physical characteristics of wireless signal attenuation, thereby achieving the highest reconstruction accuracy among all models.

\subsubsection{Dynamic Channel Gain Map}
Table~\ref{tab2} presents the quantitative results of dynamic CGM reconstruction for different models. It can be observed that the proposed model also achieves the best performance in constructing dynamic CGMs, with a larger performance gain compared to the static case. This is because, in dynamic CGMs, vehicles in addition to buildings act as obstacles, making the propagation environment more complex. By introducing physical constraints, the proposed model ensures stronger physical consistency between the generated dynamic CGMs and the actual propagation environment, thereby achieving higher reconstruction accuracy.

\begin{table}[!htbp]
\begin{center}
\caption{Quantitative comparison of dynamic CGM reconstruction results among different models}
\label{tab2}
\begin{tabular}{ m{2cm}<{\centering}  m{1.5cm}<{\centering} m{1.5cm}<{\centering} m{1.5cm}<{\centering}}
\hline
Model & NMSE & RMSE & PSNR\\
\hline
RadioUNet&0.009097&0.026797&31.760928\\
RME-GAN&0.011936&0.031072&30.333029\\ 
RadioDiff&0.007616&0.024478&32.572374\\
Proposed Model&\textbf{0.005866}&\textbf{0.022074}&\textbf{33.455685}\\
\hline
\end{tabular}
\end{center}
\end{table}

\begin{figure*}[!t]
\centering
\includegraphics[width=6in]{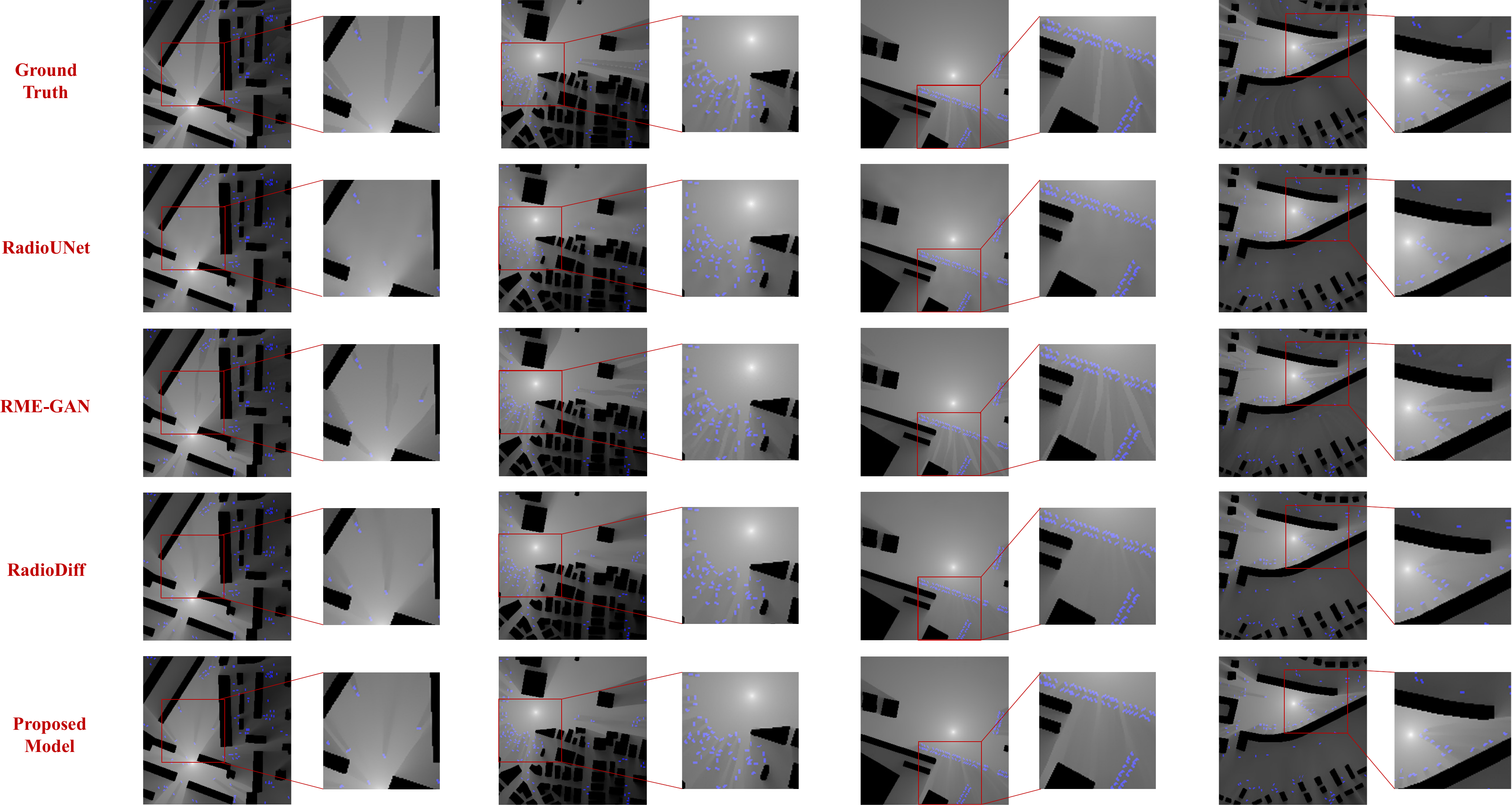}
\caption{Comparison of Dynamic CGMs Constructed by Different Models.}
\label{fig_9}
\end{figure*}

Fig. (\ref{fig_9}) shows the comparison between the dynamic CGMs reconstructed by different models and the ground-truth dynamic CGMs, where the blue dotted regions represent vehicles. As observed from Fig. (\ref{fig_9}), the dynamic CGMs reconstructed by RadioUNet still exhibit blurry edges and fail to effectively capture the partial shadowing effect caused by vehicles. Although RME-GAN produces relatively sharper edges at the locations of vehicles, it introduces large reconstruction errors in the spatial distribution of shadowed regions. RadioDiff is able to accurately reflect the shadowed areas caused by vehicles but still lacks fine-grained details. In contrast, the proposed physics-inspired model, benefiting from the introduced physical constraints, can more precisely capture the subtle path loss variations caused by small obstacles such as vehicles.

\subsubsection{Parameter Size and Construction Time}

\begin{table}[!htbp]
\begin{center}
\caption{Comparison of network parameter scales and reconstruction times among different models}
\label{tab3}
\begin{tabular}{ m{2cm}<{\centering}  m{1.5cm}<{\centering} m{1.5cm}<{\centering}}
\hline
Model & Parameter Size & Construction Time\\
\hline
RadioUNet&26.548M&0.062464s\\
RME-GAN&63.915M&0.113272s\\ 
RadioDiff&257.828M&1.367726s\\
Proposed Model&253.024M&0.675583s\\
\hline
\end{tabular}
\end{center}
\end{table}

Table~\ref{tab3} presents the comparison of network parameter scales and reconstruction times among the four CGM construction models. Since diffusion-based models involve additional components such as a VAE, a conditional encoder, and cross-attention mechanisms, they generally have larger network sizes and longer inference times. However, the proposed model achieves the highest reconstruction accuracy while maintaining a smaller number of parameters and shorter reconstruction time compared with RadioDiff. Moreover, considering that variations in CGM are mainly caused by the movement of vehicles or pedestrians, the temporal interval of such changes is typically no less than 1s. With an average reconstruction time of approximately 0.68s, the proposed model is capable of satisfying the real-time CGM construction requirement and providing high-precision CGMs for downstream communication tasks.

\subsection{Ablation study}
\begin{figure}[!t]
\centering
\includegraphics[width=6in]{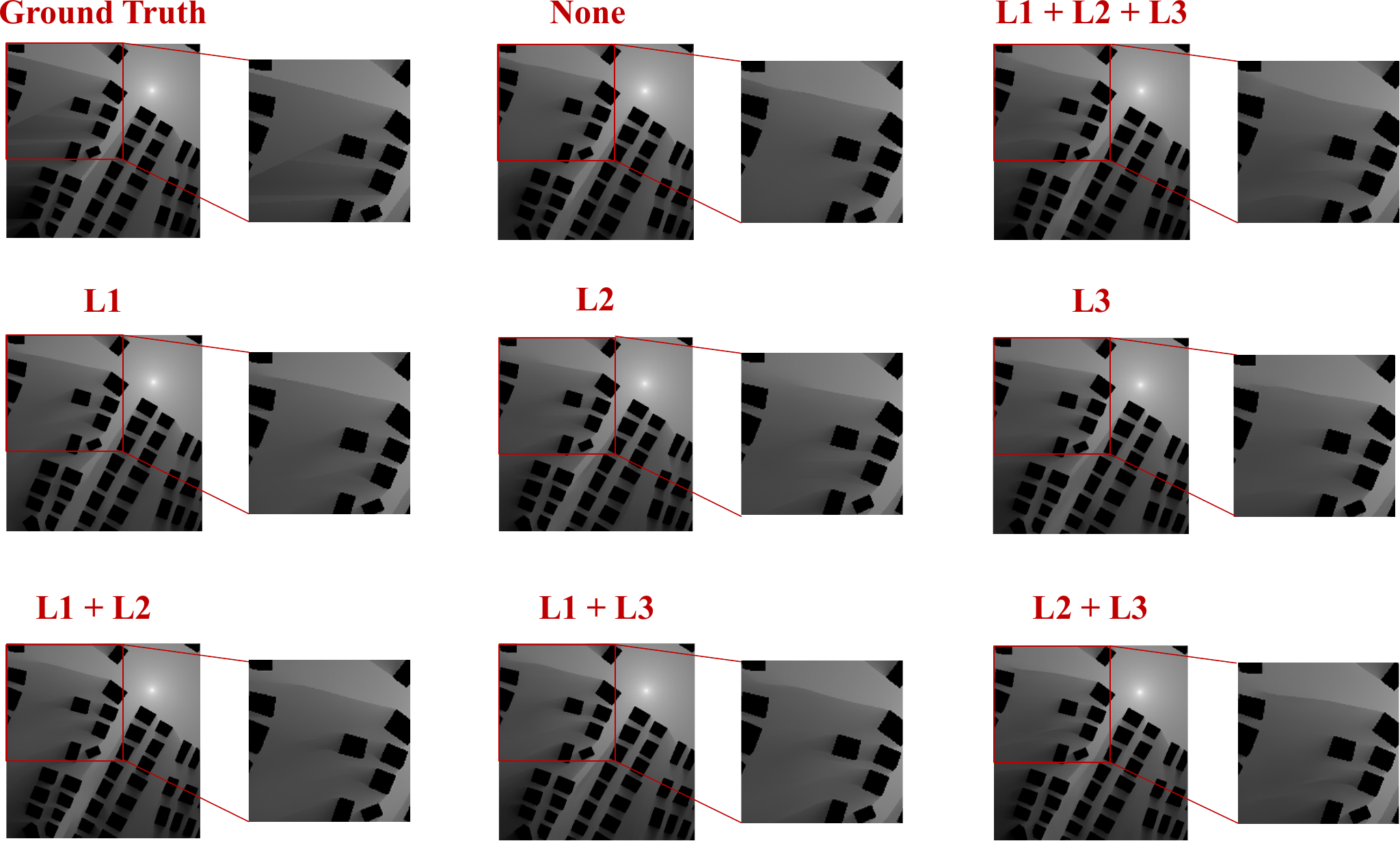}
\caption{Comparison of construction results with different physical loss terms introduced.}
\label{fig_10}
\end{figure}

In this section, we compare the experimental results obtained under four configurations: without any physics-inspired terms, with one, with two, and with all three physics-inspired terms, as shown in Table~\ref{tab4}. For simplicity, the edge loss, region propagation loss, and multi-scale feature loss are denoted as $L_1$, $L_2$ and $L_3$, respectively. "None" represents the case where no physics-inspired terms are introduced, i.e., the vanilla LDM is used for CGM construction, which yields the highest reconstruction error according to Table~\ref{tab4}.
\begin{table}[!htbp]
\begin{center}
\caption{Quantitative comparison of construction results with different physical loss terms introduced}
\label{tab4}
\begin{tabular}{ m{2cm}<{\centering}  m{1.5cm}<{\centering} m{1.5cm}<{\centering} m{1.5cm}<{\centering}}
\hline
Introduced Physical Loss Terms & NMSE & RMSE & PSNR\\
\hline
None&0.006236&0.022501&33.276135\\
$L_1$&0.004807&0.020823&33.890584\\ 
$L_2$&0.004900&0.020938&33.848756\\
$L_3$&0.004775&0.020783&33.904977\\
$L_1+L_2$&0.004884&0.020918&33.855952\\
$L_1+L_3$&0.004739&0.020738&33.921169\\
$L_2+L_3$&0.004803&0.020818&33.892383\\
$L_1+L_2+L_3$&\textbf{0.004697}&\textbf{0.020686}&\textbf{33.940059}\\
\hline
\end{tabular}
\end{center}
\end{table}

Finally, the grayscale CGMs reconstructed with different physics-inspired terms are compared in Fig. (\ref{fig_10}). By comparing them with the CGM constructed without any physics-inspired terms, the effects of the individual physics-inspired terms can be visually observed.

\section{Conclusion}
In this paper, we combine physical regularization terms with a diffusion model and propose a physics-inspired diffusion model for channel knowledge map construction. First, three physics-based regularization terms are heuristically designed according to the spatial characteristics of CGM. Then, the loss function is derived by integrating the physical regularization terms with the diffusion model objective. Finally, the training and generation architectures of the proposed physics-inspired diffusion model are developed. Experimental results demonstrate that, when only the propagation environment information is used as input, the proposed model consistently outperforms the state-of-the-art RadioDiff model, achieving performance improvements of 19.97\% and 22.98\% in static and dynamic CGM construction, respectively. Moreover, the proposed generative framework can be extended to reconstruct other types of CKM by designing additional physics-inspired terms. In future work, we plan to investigate the construction of other channel knowledge maps, such as small-scale fading feature maps and SINR maps.

\section{Appendix}
The detailed derivation of Equation \ref{deqn_ex29a} is given as follows:

\begin{equation}
\label{deqn_ex36a}
\begin{aligned}
& E_{\mathbf{x}_0 \sim q\left(\mathbf{x}_0\right)}\left[-\log p_\theta\left(\mathbf{x}_0\right)\right]
+E_{\mathbf{x}_{1:T} \sim q\left(\mathbf{x}_{1:T}\right)}\left[-\log q_R\left(\tilde{r}=0 \mid \hat{\mathbf{x}}_0\left(\mathbf{x}_{1:T}\right)\right)\right] \\
\leq &E_{q\left(\mathbf{x}_{0:T}\right)}\left[\log \frac{q\left(\mathbf{x}_{1:T} \mid \mathbf{x}_0\right)}{p_\theta\left(\mathbf{x}_{0:T}\right)}\right]
+E_{\mathbf{x}_{1:T} \sim q\left(\mathbf{x}_{1:T}\right)}\left[-\log q_R\left(\tilde{r}=0 \mid \hat{\mathbf{x}}_0\left(\mathbf{x}_{1:T}\right)\right)\right] \\
= &E_{q\left(\mathbf{x}_{0:T}\right)}\left[\log \frac{q\left(\mathbf{x}_{1: T} \mid \mathbf{x}_0\right)}{p_\theta\left(\mathbf{x}_{0:T}\right)}\right] +\int \cdots \int\left[-\log q_R\left(\tilde{r}=0 \mid \hat{\mathbf{x}}_0\left(\mathbf{x}_{1: T}\right)\right)\right] q\left(\mathbf{x}_{1:T}\right) d \mathbf{x}_{1:T} \\
= &E_{q\left(\mathbf{x}_{0:T}\right)}\left[\log \frac{q\left(\mathbf{x}_{1:T} \mid \mathbf{x}_0\right)}{p_\theta\left(\mathbf{x}_{0: T}\right)}\right] +\int \cdots \int\left[-\log q_R\left(\tilde{r}=0 \mid \hat{\mathbf{x}}_0\left(\mathbf{x}_{1: T}\right)\right)\right] \int q\left(\mathbf{x}_{0:T}\right) d \mathbf{x}_0 d \mathbf{x}_{1:T} \\
= &E_{q\left(\mathbf{x}_{0:T}\right)}\left[\log \frac{q\left(\mathbf{x}_{1: T} \mid \mathbf{x}_0\right)}{p_\theta\left(\mathbf{x}_{0: T}\right)}\right] +\int \cdots \int\left[-\log q_R\left(\tilde{r}=0 \mid \hat{\mathbf{x}}_0\left(\mathbf{x}_{1:T}\right)\right)\right] q\left(\mathbf{x}_{0:T}\right) d \mathbf{x}_{0:T} \\
= &E_{q\left(\mathbf{x}_{0:T}\right)}\left[\log \frac{q\left(\mathbf{x}_{1:T} \mid \mathbf{x}_0\right)}{p_\theta\left(\mathbf{x}_{0:T}\right)}-\log q_R\left(\tilde{r}=0 \mid \hat{x}_0\left(\mathbf{x}_{1:T}\right)\right)\right] \\
= &E_{t\sim[1,T],\mathbf{x}_{0:T}\sim q\left(\mathbf{x}_{0:T}\right)}[\frac{\tilde{A}_t^2}{2 \tilde{\beta}_t^2}\left\|\mathbf{x}_0-\mathbf{x}_\theta\left(\mathbf{x}_t, t\right)\right\|^2 +\frac{1}{2 \sigma^2}\left\|r\left(\hat{\mathbf{x}}_0\left(\mathbf{x}_t, t\right)\right)\right\|^2] \nonumber,
\end{aligned}
\end{equation}

where, the detailed proof of the following equation can be found in \cite{ho2020denoising}.
\begin{equation}
\label{deqn_ex37a}
\begin{aligned}
& E_{q\left(\mathbf{x}_{0:T}\right)}\left[\log \frac{q\left(\mathbf{x}_{1:T} \mid \mathbf{x}_0\right)}{p_\theta\left(\mathbf{x}_{0:T}\right)}\right] = E_{t\sim[1,T],\mathbf{x}_{0:T}\sim q\left(\mathbf{x}_{0:T}\right)}\left[\frac{\tilde{A}_t^2}{2 \tilde{\beta}_t^2}\left\|\mathbf{x}_0-\mathbf{x}_\theta\left(\mathbf{x}_t, t\right)\right\|^2\right] \nonumber.
\end{aligned}
\end{equation}

\bibliography{reference}
\bibliographystyle{plain}
\end{document}